\documentclass[useAMS,usenatbib]{mn2e}

\usepackage{aas_macros}	   % For shorthand in references
\usepackage{amsmath}       % Standard math type library
\usepackage{amssymb}
\usepackage{wasysym}	   % Allows the use of astronomical symbols
\usepackage{mathrsfs}      % Improves calligraphic math text
\usepackage{graphicx}      % Allows import of graphics
\usepackage{setspace}      % To manually set line spacing
\usepackage{epstopdf}      % to convert eps images to pdf images
\usepackage{verbatim}      % to allow multiline comments
\title[ADP Accretion in Major Merger Simulations]{Accretion Disc Particle Accretion in Major Merger Simulations}

\author[J. Wurster and R.J. Thacker]{J. Wurster\thanks{E-mail: jwurster@ap.smu.ca} and R. J. Thacker\\
Department of Astronomy and Physics, St Mary's University, Halifax B3H 3C3, Canada\\}

\begin{document}
\date{Accepted 2013 January 29. Received 2013 January 29; in original form 2013 January 09}
\pagerange{\pageref{firstpage}--\pageref{lastpage}} \pubyear{2013}

\maketitle

\label{firstpage}

%******************************************************************************************************************************************************************************
%Abstract
\begin{abstract}

A recent approach to simulating localized feedback from active galactic nuclei (AGN) by \citet{PNK11} uses an accretion disc particle to represent both the black hole and its accretion disc. We have extrapolated and adapted this approach to simulations of Milky Way-sized galaxy mergers containing black holes and explored the impact of the various parameters in this model as well as its resolution dependence. The two key parameters in the model are an effective accretion radius, which determines the radius within which gas particles are added to the accretion disc, and a viscous time-scale which determines how long it takes for material in the accretion disc to accrete on to the black hole itself.  We find that there is a limited range of permitted accretion radii and viscous time-scales, with unphysical results produced outside this range.  For permitted model parameters, the nuclear regions of simulations with the same resolution follow similar evolutionary paths, producing final black hole masses that are consistent within a factor of two.  When comparing the resolution dependence of the model, there is a trend towards higher resolution producing slightly lower mass black holes, but values for the two resolutions studied again agree within a factor of two.  We also compare these results to two other AGN feedback algorithms found in the literature.  While the evolution of the systems vary, most notably the intermediate total black hole mass, the final black hole masses differ by less than a factor of five amongst all of our models, and the remnants exhibit similar structural parameters.  The implication of this accretion model is that, unlike most accretion algorithms, a decoupling of the accretion rate on to the black hole and the local gas properties is permitted and obtained; this allows for black hole growth even after feedback has prevented additional accretion events on to the disc.
\end{abstract}

%******************************************************************************************************************************************************************************
% The Paper - Introduction
\section{Introduction}
\label{intro}
Observational evidence suggests that supermassive black holes exist at the centre of all galaxies with stellar spheroids (e.g. \citealp{KR95}; \citealp{FM00}).  Further evidence, such as the $M_\text{BH}$--$\sigma$ relationship (e.g. \citealp{SR98}; \citealp{FM00}; \citealp{Getal00}; \citealp{Tetal02}; \citealp{K03}; \citealp{Getal09}) and the $M_\text{BH}$--$M_\text{bulge}$ relationship (\citealp{Metal98}; \citealp{MD02}; \citealp{MH03}), suggests that the black hole and spheroid evolution are related.  Moreover, galaxies are relatively smaller today than might be naively predicted from the hierarchical model of galaxy formation, indicating that some mechanism has limited the growth of these galaxies.  One favoured explanation is that during mergers gas from the merger fuels both star formation and AGN activity (e.g. \citealp{Setal88}; \citealp{SSB05}).  The latter is likely a self-regulated process:  outflows from the black hole following a strong accretion event interact with the surrounding gas, inhibiting further accretion events, and hence limiting black hole growth (e.g. \citealp{SR98}; \citealp{F99a}; \citealp{SO04}; \citealp{K05}; \citealp{K10}).  Ultimately, the feedback from the increased AGN activity blows away some or possibly all the gas, truncating the star formation and leading to an elliptical galaxy (e.g. \citealp{SDH05l}).

Using various models, AGN feedback has been implemented in many numerical simulations (e.g. \citealt{SDH05}; \citealt{TSC06}; \citealt{ONB08}; \citealt{BS09}; \citealt{KPN09}; \citealt{DQM11}). The goals of this research have been varied, from reproducing the observed relationships between the black hole and spheroid, to other factors impacted by AGN activity, such as galaxy cluster properties (e.g. \citealp{PSS08}) or even the impact on CMB foregrounds (e.g. \citealp{STC08}).  In all these simulations, the accretion rate of gas on to the black hole is dependent on extrapolating the macroscopic gas properties to the microscopic scale around the black hole.  This is true for viscous accretion (e.g. \citealp{DQM11}; \citealp{HQ11}), drag accretion \citep{ONB08}, and the numerical form of Bondi accretion rate \citep{B52},  
\begin{equation}
\label{mBondi}
\dot{M}_\text{Bondi} = \frac{4\pi\alpha G^2 M^2_\text{BH} \rho}{\left(c_\text{s}^2 + v_\text{rel}^2\right)^{3/2}},
\end{equation}
where $\alpha$ is a numerical parameter used to account for the limited dynamic range in the simulations, $\rho$ and $c_\text{s}$ are the gas density and sound speed around the black hole, $v_\text{rel}$ is the relative velocity between the gas and the black hole, and $M_\text{BH}$ is the mass of the black hole.  This accretion rate has been used in merger (e.g. \citealt{SDH05}; \citealt{DSH05}) and in cosmological (e.g. \citealp{SSDH07}; \citealp{BS09}) simulations.

Most algorithms also limit the mass accretion to be no greater than the Eddington accretion rate, 
\begin{equation}
\label{mEdd}
\dot{M}_\text{Edd} \equiv \frac{4 \pi G M_\text{BH} m_\text{p}}{\epsilon_\text{r} \sigma_\text{T} c},
\end{equation}
where $m_\text{p}$ is the proton mass, $\sigma_\text{T}$ is the Thomson cross section, and $\epsilon_\text{r}$ is the radiative efficiency (i.e. the mass-to-energy conversion efficiency); we set $\epsilon_\text{r} = 0.1$ \citep{SS73}.  This rate differs from the previously considered accretion rates in that it depends only on the black hole mass, and not the local gas properties.  

One of the assumptions frequently used in these models is that any accreted gas is immediately transferred to the black hole. This is clearly an unrealistic assumption as material must first shed its angular momentum. 
Indeed some models argue that understanding this process is one of the key factors in modelling AGN feedback; for example, the viscous accretion rate in \citet{DQM11} accounts for angular momentum of the gas. However, the gas in this model is nonetheless instantly accreted on to the black hole. In reality the gas would be expected to settle on to a circular orbit of radius $R_\text{circ}$, which is set by the angular momentum of the gas and whatever processes cause it to shed angular momentum while collapsing (cf. \citealp{HNPK11}), resulting in an accretion disc around the black hole.  The gas with the lowest angular momentum would then travel through the disc and eventually be accreted on to the black hole (e.g. \citealp{K10}).

Motivated by these ideas, \citet{PNK11} (hereafter PNK11) have implemented a two-stage accretion algorithm using an accretion disc particle (ADP), which models both the black hole and accretion disc.  In the first stage, nearby gas, the bulk of which will have shed large amounts of angular momentum to reach this radius, is accreted on to the accretion disc; this accretion rate is dependent only on the relative positions of the black hole and the gas particle.  In the second stage, the gas is accreted from the accretion disc on to the black hole.  This method incorporates the delay between the time gas is accreted on to the disc and when it is finally accreted on to the black hole, and at the same time decouples the black hole's accretion rate from the instantaneous gas properties around the black hole.  

While the ADP model implemented in \citet{PNK11} relies upon resolving scales far smaller than those that can be resolved in merger simulations, the two stage approach, and perhaps more specifically the incorporation of a delay period before accreting on to the black hole, is an issue worthy of investigation in merger simulations which can have time-steps in the few thousand year range. But this consideration highlights the fact that resolution dependence must be considered.  For example, at very low mass resolution and the associated low time resolution, concerns about accretion delays are likely less significant as the ratio of time-step and delay time get closer to unity.  Therefore investigating the precise resolution dependence is important.  We also emphasize that the ADP model needs additional features to be implemented in a merger simulation, but we have drawn on prior work and use methods that have been well studied in the literature.

The layout of this paper is as follows: In section \ref{sims} we discuss salient details of the simulation including the PNK11 model and how we have augmented it to allow for black hole tracking. In section \ref{results} we compare the behaviour of the merger simulations and the sensitivity of final state diagnostics, specifically the black hole mass, as a function of the model parameters. We end with a brief review.

%******************************************************************************************************************************************************************************
%The paper - Numerical Simulations
\section{Numerical Simulations}
\label{sims}

We test the ADP method in a simulation of a major merger of two Milky Way-sized galaxies.  Each galaxy contains a dark matter halo, hot gas halo, stellar disc, gas disc, a stellar bulge, and a black hole.  The galaxies are initially separated by 70 kpc, and placed on a parabolic trajectory around one another similar to the merger considered in \citet{SDH05}. The construction of the galaxies and definitions of the structural parameters are given in Appendix \ref{simsGM}, while the parameters of each component are presented in Table \ref{galprops}. The masses and particle numbers used are summarized in Table \ref{breakdown}.

The simulations were run using the parallel version of {\sc Hydra} (\citealp{CTP95}; \citealp{TC06}), which uses an Adaptive Particle-Particle, Particle-Mesh algorithm \citep{C91} to calculate gravitational forces and the Smooth Particle Hydrodynamics method (SPH; \citealp{GM77}; \citealp{L77}) to calculate gas forces.  The star formation algorithm is described in \citet{TC00}.

To implement an AGN feedback algorithm motivated by the PNK11 model, we have had to make some modifications and additions. We have modified how feedback is returned to the local gas, and added both a black hole advection algorithm and a black hole merger algorithm. None of the issues related to black hole advection or mergers are addressed within the PNK11 paper since they consider accretion of gas on to a black hole that is embedded within an initially spherical gas cloud.
\begin{center}
\begin{table*}
\begin{minipage}{\textwidth}
{\small
\hfill{}
\begin{tabular}{l|r r r r r r}
    \hline
  &                           & \multicolumn{2}{c}{Fiducial resolution} &&  \multicolumn{2}{c}{Low resolution} \\
\cline{3-4}
\cline{6-7}
 & \multicolumn{1}{c}{Total mass}                 & \multicolumn{1}{c}{Particle mass}           & \multicolumn{1}{c}{Number of}  && \multicolumn{1}{c}{Particle mass}           &  \multicolumn{1}{c}{Number of}\\
 & \multicolumn{1}{c}{(10$^{10}$ M$_{\astrosun}$)}& \multicolumn{1}{c}{(10$^5$ M$_{\astrosun}$)}& \multicolumn{1}{c}{particles}  && \multicolumn{1}{c}{(10$^5$ M$_{\astrosun}$)}&  \multicolumn{1}{c}{particles}\\
\hline
\hline
Dark matter halo & 89.92    & 11.75 &   765 000 && 89.92 & 100 000 \\
Hot gas halo     &  0.60    &  0.36 &   165 343 &&  2.77 &  21 619 \\
Stellar bulge    &  1.34    &  2.37 &    56 649 && 18.10 &   7 407 \\
Stellar disc     &  3.56    &  2.37 &   150 375 && 18.10 &  19 662 \\
Gas disc         &  0.54    &  0.36 &   150 375 &&  2.77 &  19 662 \\
Black hole       &10$^{-5}$ &  1.00 &         1 &&  1.00 &       1 \\
\hline
Total            & 95.96    &       & 1 287 743 &&       & 168 351 \\
\hline
\end{tabular}}
\hfill{}
\caption{Component breakdown for each galaxy.  The accretion disc initially has zero mass, thus the initial mass of the ADP is the same as the black hole.}
\label{breakdown} 
\end{minipage}
\end{table*}
\end{center}

%-----------------------------------------------------------------------
\subsection{Implementing the PNK11 model: Approach to accretion}
\label{IPacc}
The black hole accretion algorithm is the same as in PNK11.  In this method the accretion disc particle (ADP) is a collisionless sink particle containing both the black hole and its tightly bound accretion disc.  The mass of the ADP is given by
\begin{equation}
\label{adapmass}
M_\text{ADP} = M_\text{BH} + M_\text{disc},
\end{equation}
where $M_\text{BH}$ is the mass of the black hole and $M_\text{disc}$ is the mass of the accretion disc. As in PNK11, we initialise $M_\text{disc} = 0$, while we choose the initial black hole mass in each galaxy to be 10$^{5}$ M$_{\astrosun}$.  The ADP has an associated `smoothing length', $h_\text{ADP}$, which is calculated at every iteration by $h_\text{ADP} = \max \left( h_\text{ADP}, h_\text{min} \right)$, where a sphere with radius $2h_\text{ADP}$ around the ADP includes 60 gas particles, and $h_\text{min}$ is the smallest resolved smoothing length in the SPH solver.  The smoothing length is used to calculate the gas properties at the location of the ADP, which is used for analysis and returning feedback energy.  The ADP is also given an accretion radius, $R_\text{acc}$, which is a free parameter of the model that determines at what radii particles are considered to have accreted on to the accretion disc. Nominally, and indeed in PNK11, this radius is expected to be on the order of a pc. While merger simulations can reach these small radii by considering arbitrarily small values of  $R_\text{acc}$, it is worth emphasizing that the properties of the gas on these scales is not resolved. Indeed this is a generic problem with feedback models in general, namely that they use input parameters at the edge of model resolution. However, since this is an unavoidable problem, we continue on with this issue noted and examine the impact of varying the value of $R_\text{acc}$ by up to a factor of 10.

Whenever a gas particle comes within $R_\text{acc}$ of the ADP, it is instantly accreted on to the accretion disc, thus $M_\text{disc} \rightarrow M_\text{disc} + m_\text{gas}$, where $m_\text{gas}$ is the mass of the gas particle.  The accreted particle's mass and momentum are added to the ADP, and then the accreted particle is removed from the simulation.  This capture rate is not limited in any way, and solely depends on the particles' relative locations.  Once accreted on to the accretion disc, it takes time for the gas to be transported through the disc so that it can finally be accreted on to the black hole.  This time delay is on order of the disc viscous time, $t_\text{visc}$, which must be larger than the dynamical time at the accretion radius; PNK11 set $t_\text{visc}$ as a free parameter, but argue that it should be $t_\text{visc} \sim $10--100 Myr in galaxy formation simulations.  

The mass accretion rate on to the black hole from the accretion disc is given by,
\begin{equation}
\label{acc}
\dot{M}_\text{BH} = \min \left( \frac{M_\text{disc}}{t_\text{visc}}, \dot{M}_\text{Edd} \right).
\end{equation}
To preserve mass continuity, the mass added to the black hole is removed from the accretion disc although the overall ADP mass remains the same.  As in most numerical implementations of AGN feedback, the accretion rate on to the black hole is Eddington-limited; this moderates the growth of the black hole based upon a physical limit rather than just from the numerical accretion on to the disc.

%-----------------------------------------------------------------------
\subsection{Implementing the PNK11 model: Changes and additions for merger simulations}
\label{IPadd}
\subsubsection{Feedback}
\label{Feedback}
PNK11 use the same feedback scheme as described in \citet{NP10}.  In this scheme, feedback energy is returned by adding wind particles which are radially directed away from the black hole; the wind particles have momentum $p_\text{wind} = 0.1m_\text{gas} \sigma$, where $\sigma$ is the velocity dispersion of the host galaxy.  In our simulation, it is impractical to continually add wind particles since the particle load in the solver will climb rapidly, so we instead follow the wind prescription used in \citet{DQM11}.  Here, the feedback rate is given by
\begin{equation}
\label{pfeed}
\dot{p} = \tau \frac{L}{c},
\end{equation}
where $\tau = 10$ is the assumed infrared optical depth, and $L = \epsilon_\text{r} \dot{M}_\text{BH} c^2$ is the luminosity, where $\epsilon_r = 0.1$ is the radiative efficiency.  The momentum is returned radially, such that every gas particle within the ADP's radius of influence, $r_\text{inf} \equiv 2h_\text{ADP}$, receives an equal acceleration.  We reiterate that $r_\text{inf}$ has no explicit dependence on $R_\text{acc}$ and is being recalculated at every iteration.
 
\subsubsection{Black hole advection and mergers}
The black hole advection algorithm is a modified version of that presented in \citet{ONB08} and used in Model WT in \citet{WT13}.  Here, the ADP is displaced towards the centre of mass of the sphere with radius $r_\text{inf}$ which is centred on the ADP.  The distance it is displaced is
\begin{equation}
\label{deltal}
\Delta l = \min(0.10h_\text{ADP}, 0.30\left| {\bf v} \right| \text{d}t, d_\text{CM}),
\end{equation}
where ${\bf v}$ is the velocity of the black hole, $\text{d}t$ is the time-step, and $d_\text{CM}$ is the distance from the black hole to the centre of mass.  The coefficients are based upon those in \citet{ONB08}, but modified for our higher resolution.  Following the approach of \citet{ONB08}, we have selected the parameters to suppress the oscillations of the black hole particles such that the maximum amplitude of the black hole oscillation between the start of the simulation and first periapsis is less than 0.25 per cent of the core radius.  This method continually displaces the black hole towards the centre of mass to counter any two-body interactions that may try to displace the black hole from the bottom of the potential well; limiting the distance preserves the possibility of the black hole remaining in a gas void.  One option we considered but did not implement was to couple the ADP to the gas particle that has the lowest potential energy, is within $r_\text{inf}$ of the ADP and satisfies $v_\text{rel} < 0.25c_\text{s}$, where $v_\text{rel}$ is the relative velocity between the ADP and the gas particle and $c_\text{s}$ is the local sound speed; generally, this can lead to large artificial displacements and prohibits the black hole from remaining in a gas void, but with this specific accretion algorithm, it would artificially increase the accretion rate by arbitrarily moving the ADP within $R_\text{acc}$ of a gas particle.  A second option we considered was to use a tracer particle whose mass is $\sim$100--1000 times greater than any other particle; although this produces a smooth black hole trajectory, the additional mass affects the evolution time when compared to non-tracer particle models.

Finally, when the ADP's pass within each others smoothing lengths and have a relative velocity less than the local sound speed, they are assumed to instantly merge.  The merged black hole has the combined mass of the progenitor black holes, and the merged accretion disc has the combined mass of the progenitor discs, thus all masses are conserved.  This is similar to the numerical merger procedure used in \citet{SDH05}.

%-----------------------------------------------------------------------
\subsection{Parameter space and resolution dependence}
\begin{figure}
\begin{center}
\includegraphics[width=1.0\columnwidth]{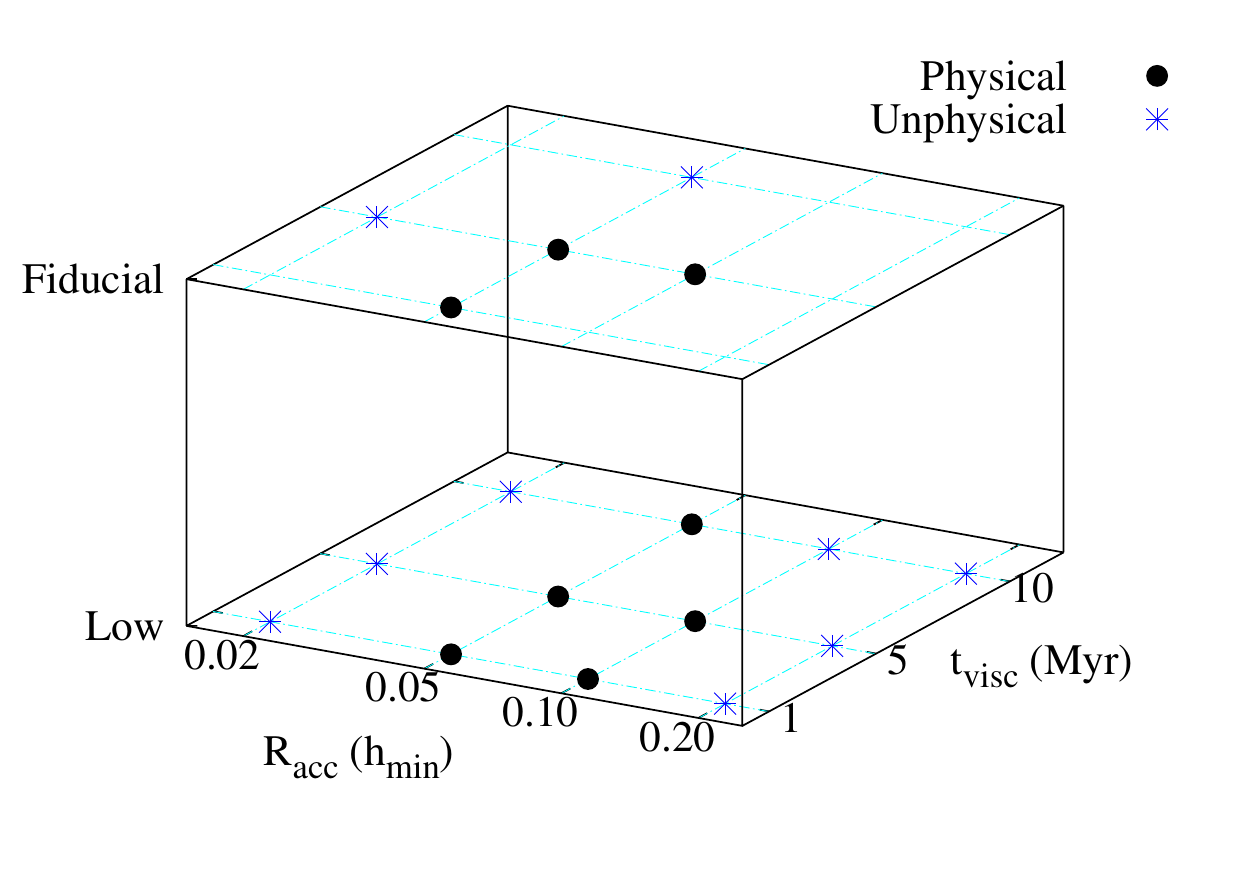}
\end{center}
\caption{The parameter space that we tested.  Each point corresponds to a model; the physical/unphysical designation will be discussed in \S \ref{results})}\label{pspace}  
\end{figure}

To understand the impact of the two free parameters, $R_\text{acc}$ and $t_\text{visc}$, we ran a suite of simulations within this parameter space and considered two separate resolutions for a total of 17 simulations.  We plot where the simulations lie in the parameter space in Fig. \ref{pspace}, where each point corresponds to a model.  Our parameter space is 
\begin{enumerate}
\item $\text{resolution} \in \{\text{low}, \text{fiducial}\}$,
\item $R_\text{acc}/h_\text{min} \in \{ 0.02, 0.05, 0.10, 0.20\}$, and 
\item $t_\text{visc}/\text{Myr} \in \{ 1, 5, 10 \}$.  
\end{enumerate}
In Table \ref{Racc}, we convert the accretion radius to physical units for both resolutions.  Our model naming convention is understood as follows: `PNK$_{resolution}$r$\left( 100R_\text{acc}/h_\text{min}\right)$t$\left( t_\text{visc}/\text{Myr}\right)$'.  If we do not include a resolution when referring to a model, then we are referring to both resolutions.  In terms of wall-clock time, despite having a modest particle content, the fiducial resolution models still take over a month to run due to the large number of time steps required, the lack of multiple time-steps in the {\sc Hydra} solver and also an $\mathcal{O}(n^2)$ slowdown that occurs as large numbers of SPH particles reach the minimum smoothing length of the solver.  We thus have a limited number of fiducial resolution simulations.  

\begin{center}
\begin{table}
{\small
\hfill{}
\begin{tabular}{l|c c }
    \hline
$R_\text{acc}$   & Fiducial resolution  &  Low resolution \\
($h_\text{min}$) &  (pc)                &  (pc)		  \\
\hline
\hline
0.02		 &  0.731		&  1.827  \\
0.05		 &  1.827		&  4.567  \\
0.10		 &  3.654		&  9.133  \\
0.20		 &  Not run		& 18.266  \\
\hline
\end{tabular}}
\hfill{}
\caption{The accretion radii, $R_\text{acc}$, in our parameter space, given in simulation and physical units.}  \label{Racc} 
\end{table}
\end{center}

%******************************************************************************************************************************************************************************
%The paper - Results
\section{Results}
\label{results}
\subsection{General evolution}
\begin{figure*}
\begin{center}
\includegraphics[width=1.0\textwidth]{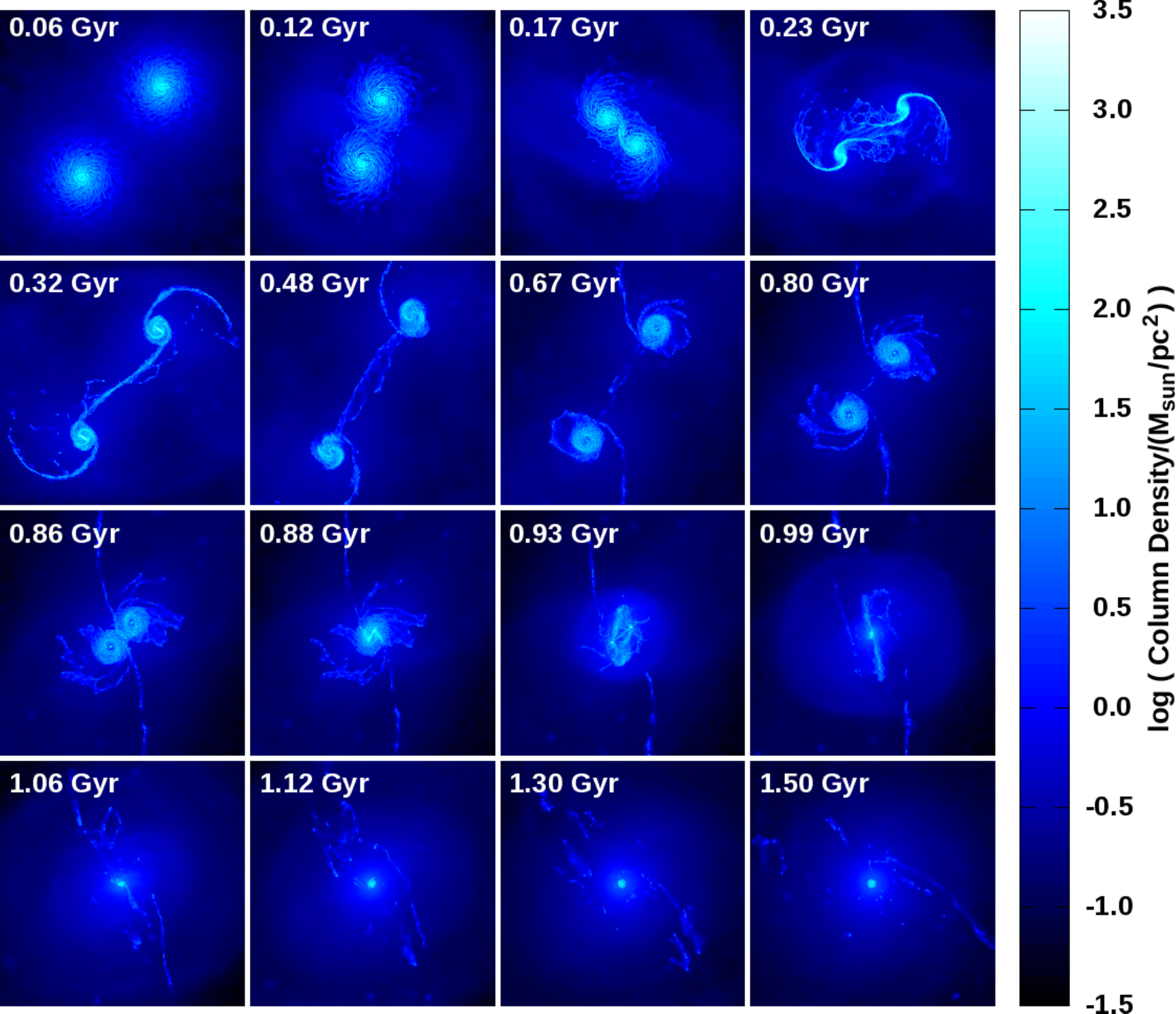}
\end{center}
\caption{Evolution of gas column density for Model PNK$_\text{f}$r05t05.  Times from the onset of the simulation are listed in each frame; each frame measures 100 kpc per side, with an image resolution of 98 pc/pixel.}\label{r_snapR}  
\end{figure*}

Each of our models was evolved for 1.5 Gyr through a merger event, similar to that of \citet{SDH05}.  We found that each model followed a similar qualitative history; in Fig. \ref{r_snapR}, we plot the evolution of the gas column density of Model PNK$_\text{f}$r05t05. 
From this evolution, we see four significant epochs: first periapsis, apoapsis, second periapsis and core merger, occurring at $\sim$170, 480, 880 and 990 Myr, respectively; the low resolution models reach second periapsis and core merger 25 and 6 Myr, respectively, earlier than their fiducial resolution counterparts.   In all our fiducial resolution models, the black holes merge at 1.01 Gyr, and their low resolution counterparts merge 10--60 Myr later.  The later merger in the low resolution models is a result of the binary black holes oscillating about one another with a greater amplitude than in the fiducial resolution models; this is from both the low resolution black holes being more massive by this epoch (see section \ref{rs}), and there being fewer particles to induce drag on the black hole system.

The final remnant is a reformed gas disc and flattened stellar ellipsoid surrounded by a hot gas halo; the remnant is further discussed in section \ref{frms}.  The final gas discs in the fiducial resolution models have scale lengths of approximately 0.5 kpc, compared to the initial scale length of 2.46 kpc and are essentially an order of magnitude less massive than the discs in the initial conditions.  Of the gas that was initially in a disc, 89 per cent of it is either accreted on to the black hole or converted into stars.  With a final star formation rate of less than 0.5 M$_{\astrosun}$/yr, the final configuration is a red and dead elliptical, as expected \citep{SDH05l}.

\begin{figure}
\begin{center}
\includegraphics[width=1.0\columnwidth]{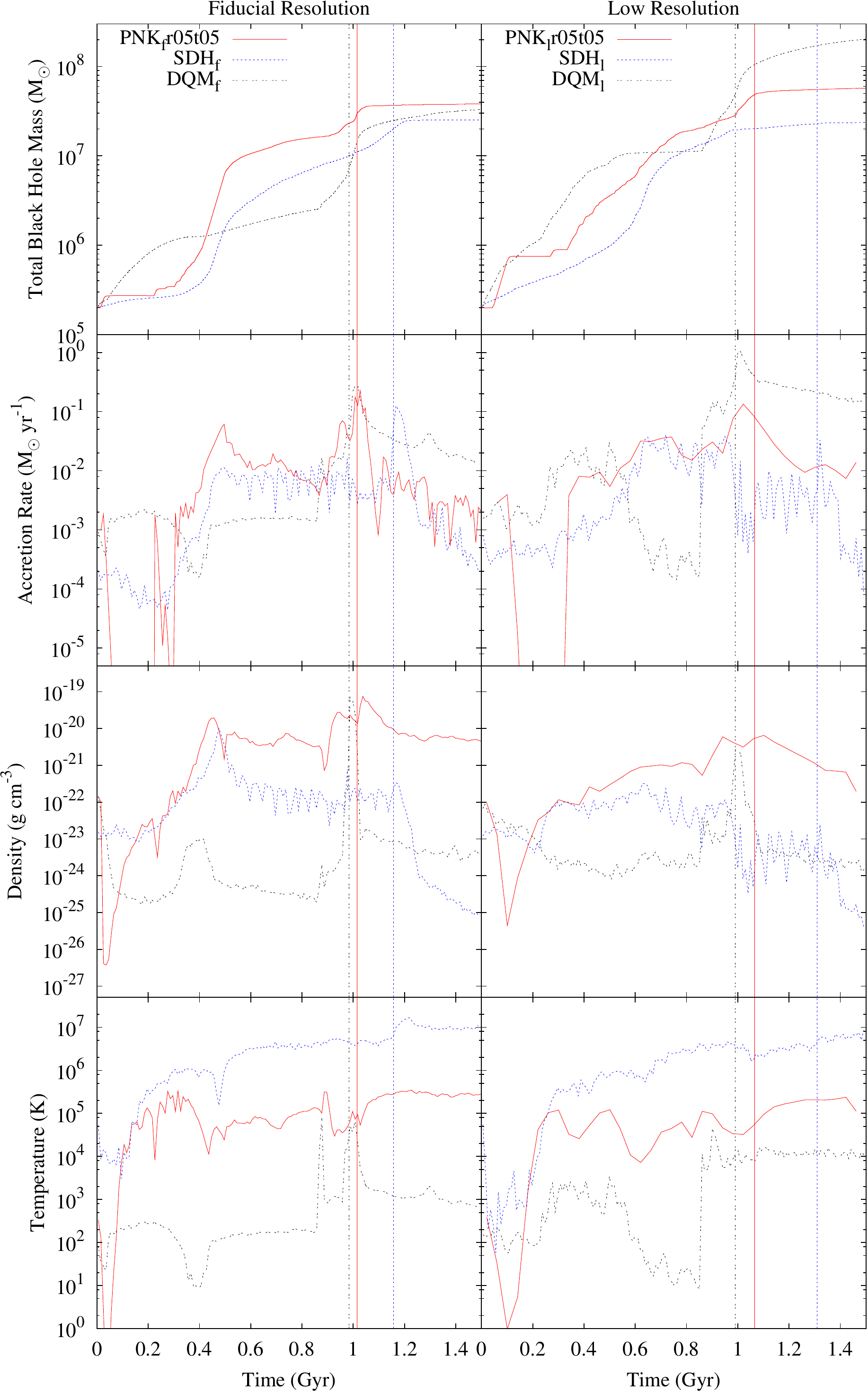}
\end{center}
\caption{\emph{Top to bottom}: Total black hole mass, accretion rates on to the black hole, gas density and gas temperature within $r_\text{inf}$ of the ADP/black hole particle.  The bottom three rows are geometrically averaged over both black holes in bins of 10 Myr.  The left panels are from the fiducial resolution simulations and the right panels are from their low resolution counterparts.   The vertical line of the same linestyle indicates when the black holes merge.} \label{r_prop_ge}   
\end{figure}

To quantify this evolution, we have plotted the total black hole mass, the accretion rates on to the black hole, and the gas density and gas temperature within $r_\text{inf}$ of the ADP for Model PNKr05r05 in Fig. \ref{r_prop_ge}; the accretion rates and gas properties are geometrically averaged over both black holes and plotted in bins of 10 Myr.

For the first $\sim$300 Myr, $r_\text{inf} > 2h_\text{min}$, thus there are only 60 gas particles within this radius; after $\sim$300 Myr, $r_\text{inf} = 2h_\text{min}$, and there are several thousand particles within this radius.  There is an increasing accretion rate after first periapsis, which peaks near apoapsis; there is a second peak at core merger.  The gas density in the core also increases from first periapsis to apoapsis, and again at core merger.  The core temperature stays relatively moderated after first periapsis.  The initial decline in gas density and temperature is a result of the feedback energy carving out a void around the black hole in the initial density field which is quite symmetric.  The SPH particles that contribute to the density and temperature at the location of the black hole are at $r \sim 0.9r_\text{inf}$, thus provide only a minimal contribution.  Thus, these low densities and temperatures are essentially a result of the smoothing kernel breaking down due to a poor distribution of the particles within 2$h_\text{ADP}$.  At later times the number of particles increases and symmetry is lost 
which prevents any spurious temperature and density values from being calculated.

%-----------------------------------------------------------------------
\subsection{Comparison to other models in the literature}
We briefly compare Model PNKr05t05 to two other models found in the literature: Model SDH, which is designed to reproduce the algorithm used in \citet{SDH05} and Model DQM, which is designed to reproduce the algorithm used in \citet{DQM11}.  Model SDH uses the Bondi accretion rate given in \eqref{mBondi}, where $\alpha = 100$, to calculate the rate of gas accretion on to the black hole; half a per cent of the accreted mass is converted in to feedback energy which is returned thermally by adding energy to the nearby gas particles, following the smoothing kernel.  Model DQM uses a viscous accretion rate, 
\begin{equation}
\label{Visc}
\dot{M}_\text{visc} = 3 \pi \delta \Sigma \frac{c_\text{s}^2}{\Omega}, 
\end{equation}
where $\delta = 0.05$ is the dimensionless viscosity, $\Sigma$ is the mean gas surface density, and $\Omega = \sqrt{GM/r_\text{inf}^3}$ is the rotational angular velocity of the gas, which is based upon the multi-scale SPH simulations by \cite{HQ10}; the feedback algorithm is same as in our PNK models.   We plot the total black hole mass, the accretion rates on to the black hole, and the gas density and gas temperature within $r_\text{inf}$ of the black hole for Models SDH and DQM in Fig. \ref{r_prop_ge}. 

We briefly compare Model PNKr05t05 to two other models found in the literature: Model SDH, which is designed to reproduce the algorithm used in \citet{SDH05} and Model DQM, which is designed to reproduce the algorithm used in \citet{DQM11}.  Model SDH uses the Bondi accretion rate given in \eqref{mBondi} to calculate the rate of gas accretion on to the black hole; the feedback energy is returned thermally by adding energy to the nearby gas particles.  Model DQM uses a viscous accretion rate, which is based upon the multi-scale SPH simulations by \cite{HQ10}; the feedback algorithm is same as in our PNK models.   We plot the total black hole mass, the accretion rates on to the black hole, and the gas density and gas temperature within $r_\text{inf}$ of the black hole for Models SDH and DQM in Fig. \ref{r_prop_ge}. 
  
Although all three accretion and feedback algorithms make different physical assumptions, the final total black hole masses in the fiducial (low) resolutions are equal within a factor of 1.52 (4.62).  The accretion histories, however, vary considerably.  As expected, Models SDH and DQM have continual accretion which starts immediately, but the accretion rate in PNKr05t05 is punctuated by periods of nearly zero accretion, especially at early times.  The latter is a result of feedback from the first few accretion events creating a small and transient void around the ADP, as can be seen in the core density profile.  After first periapsis, tidal torques are large enough and feedback energy is low enough to allow additional accretion events.

The core density of Model DQM is the lowest of these three models; in Model DQM$_\text{f}$, kinetic feedback builds up in the gas, and after 60 Myr, a void of $\sim 0.55r_\text{inf}$ is formed and maintained for the duration of the simulation.  Although Model PNKr05t05 uses the same feedback algorithm, the initially lower accretion rate prevents this build up of kinetic feedback, thus this void never forms, hence the higher core density.  The core temperature is highest for Model SDH, where thermal feedback energy directly increases the temperature of the gas.  The core heating of Models PNK and DQM is from shock heating and star formation; the higher accretion and star formation (not shown) rates in PNKr05t05 after first periapsis results in more heating than in Model DQM.

%-----------------------------------------------------------------------
\subsection{Acceptable parameter ranges}
\label{pur}

We ran 12 low resolution simulations to probe our entire $R_\text{acc}$--$t_\text{visc}$ parameter space. At our fiducial resolution we only tested five models due the time required to complete individual simulations.  After analysing the results, each simulation was classified as either physical or unphysical, as is indicated by the dot type in Fig. \ref{pspace}. The basis for our definition of physical or unphysical relies upon a combination of structural evolution and how close the final remnant lies to the $M_\text{BH}$--$\sigma$ relationship. We note that relying upon the $M_\text{BH}$--$\sigma$ relationship is perhaps not a strong constraint because the masses of the seed black holes could be changed, however it remains a commonly used approach in these simulations and we thus proceed cautiously with this issue noted.

For $R_\text{acc}$ values that are very small the accretion rate on to the black hole will be limited. The expectation in this case is that the resulting remnant will fall below the $M_\text{BH}$--$\sigma$ relation. Our numerical experiments show this to be true for all the models with $R_\text{acc} = 0.02h_\text{min}$, with the black hole mass in PNK$_\text{f}$r02t05 being 4.5 times lower than expected, assuming the velocity dispersion remains constant.  These models also suffer from sensitivity to operation ordering in the solver.  At a given resolution, we would expect the absorption of particles to occur at the same time provided that the operations in the solver are executed in the same order, and hence produce the same round-off. However, when running in parallel with dynamic load-balancing this is no longer the case and round-off in accumulations can produce subtle differences around or just above machine precision.  Thus, with too small of an accretion radius, pseudo-random issues due to machine precision dominate the accretion, and the results of a given simulation are not easily reproducible.  On the basis of lying beneath the $M_\text{BH}$--$\sigma$ relation and the exhibited numerical sensitivity, we classify all of our $R_\text{acc} = 0.02h_\text{min}$ models as `unphysical'.

\begin{figure}
\begin{center}
\includegraphics[width=1.0\columnwidth]{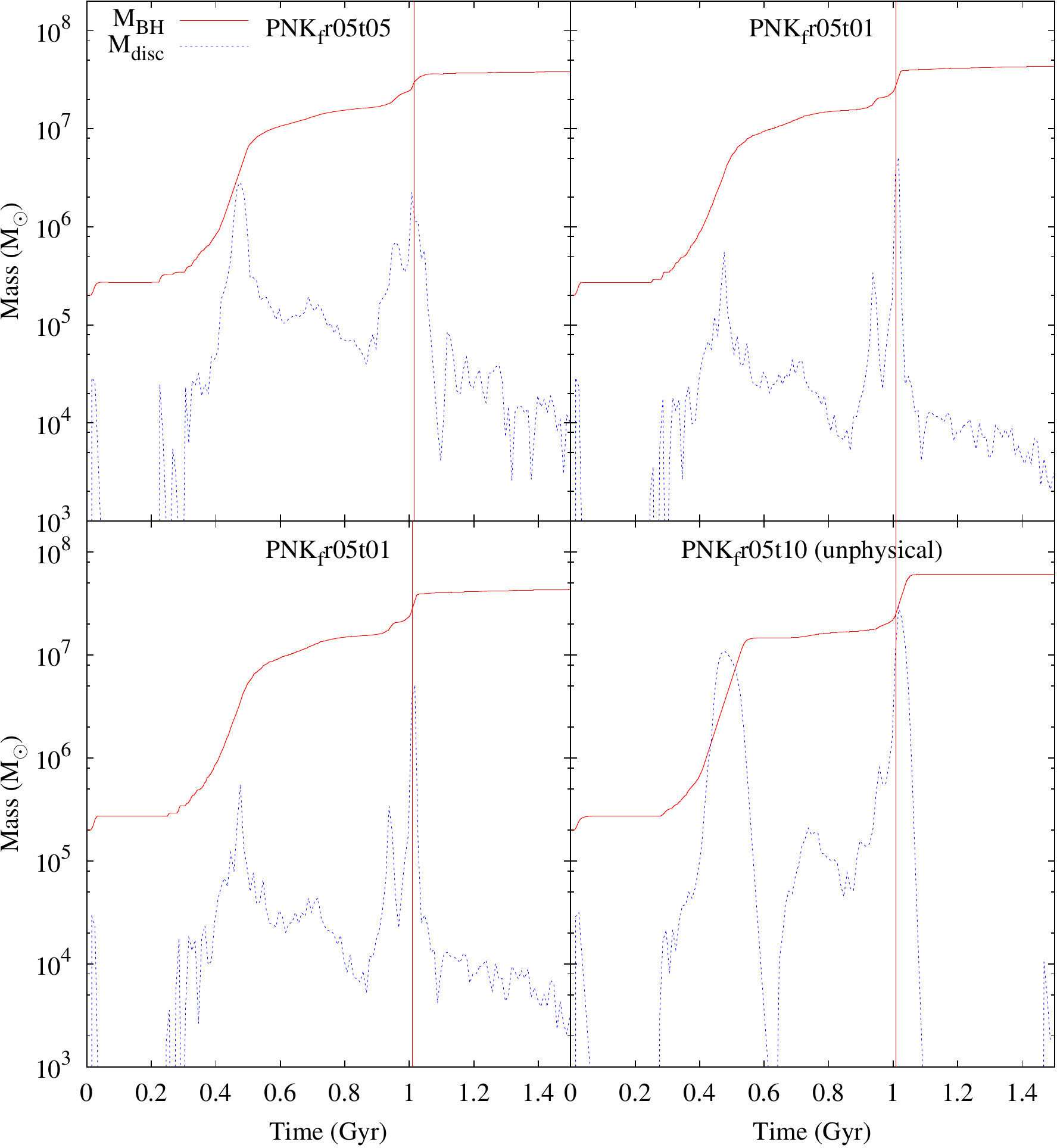}
\end{center}
\caption{Total black hole mass (solid red line) and total accretion disc mass (dashed blue line; geometrically averaged over both black holes and plotted in bins of 10 Myr) for four fiducial resolution models.  The vertical line indicates when the black holes merge.}\label{r_masses}  
\end{figure}

At the other extreme of our parameter space, there can be unphysical results if the accretion rate on to the disc is too high.  If $R_\text{acc}$ is too large, then many particles can pass within this radius and be accreted on to the disc, regardless of the amount of feedback.  Likewise, a large $t_\text{visc}$ means that the feedback energy is being returned at a lower rate, thus particles have less of an obstacle to overcome to pass within $R_\text{acc}$ when influenced by outside forces (i.e. tidal torques from the interacting galaxies).  Both result in a large accretion rate on to the disc, where the accretion disc mass can surpass the mass of the black hole (see bottom right panel of Fig. \ref{r_masses}; the remaining three panels show the profiles for physical models, where the peak accretion disc mass is $\sim5\times 10^6$M$_{\astrosun} < M_\text{BH}$).  
This unreasonably large accretion rate on to the disc yields a large and continual accretion rate from the disc on to the black hole, which results in a vast amount of feedback energy.  Once a critical amount of kinetic feedback is injected into the gas, namely that sufficient to overcome the nuclear gravitational binding energy, a large, unphysical void is formed in the galaxy. This in turn suppresses the accretion rate on to the ADP, but not from the disc on to the black hole.  The accretion disc is ultimately depleted, the feedback ends and the void recollapses to begin another cycle. In situations where a large radius and large viscous time-scale are included it is possible to remove the gas from the system.

These unphysical voids often foreshadow the total disruption of the remnant.  Starting at 1.04 Gyr in Model PNK$_\text{f}$r05t10, there is a catastrophic explosion, and the system is totally disrupted within 60 Myr.  In other models, specifically our low resolution models, the core merger causes an increase in the accretion rate, leading to a final feedback episode that disrupts the system, although often less catastrophically than Model PNK$_\text{f}$r05t10.  While the system is highly disrupted, there are often no noticeable voids, and the process is `gentle' enough that the gas begins to recondense into a disc and cloud.  We cautiously and liberally classify these results as physical.

\begin{figure}
\begin{center}
\includegraphics[width=1.0\columnwidth]{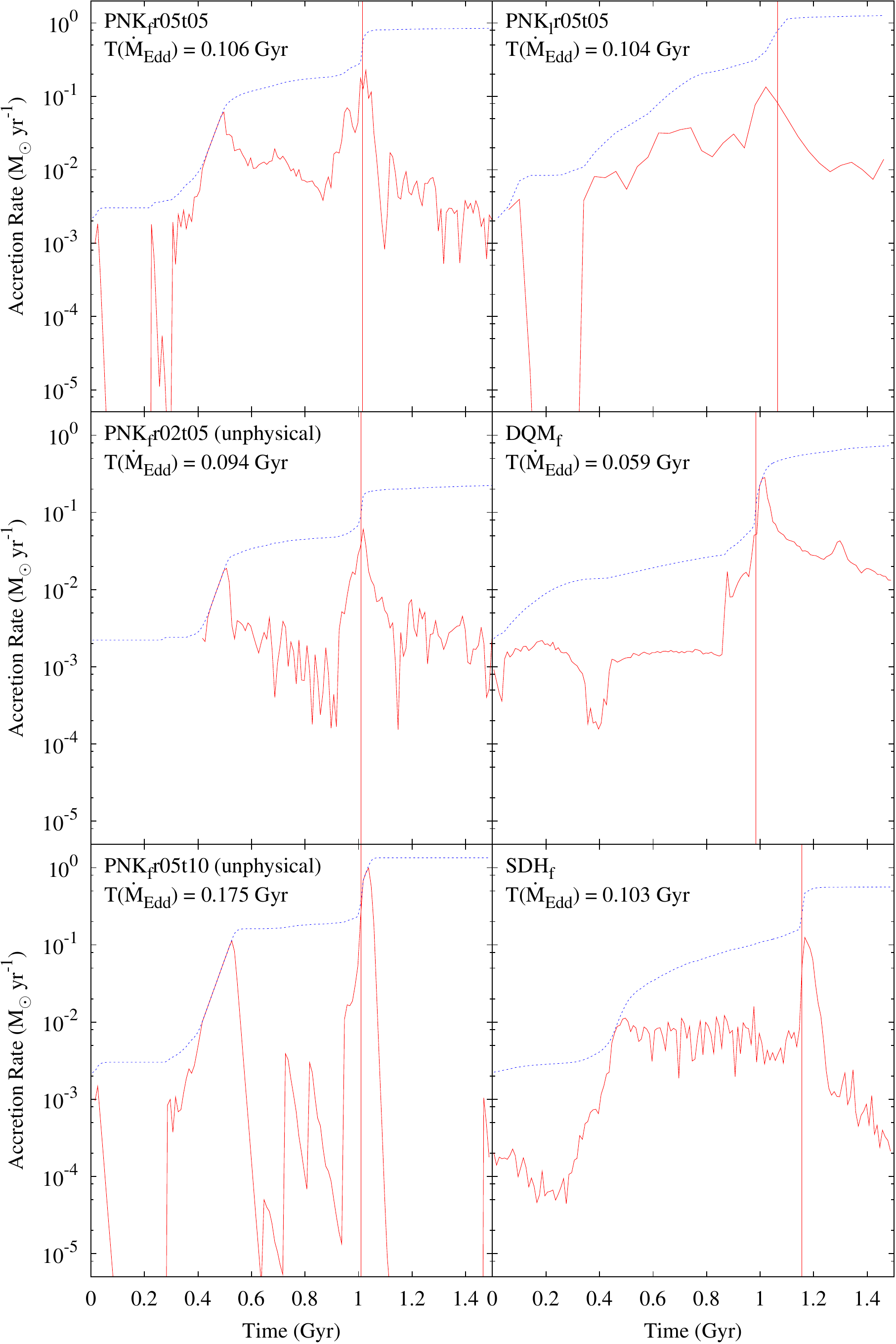}
\end{center}
\caption{Actual (solid red line) and Eddington (dashed blue line) accretion rates for six models, geometrically averaged over both black holes and plotted in bins of 10 Myr.  The vertical line indicates when the black holes merge.  Model names and length of time the system is undergoing Eddington accretion are listed in the top left of each panel.}\label{r_accEL}  
\end{figure}

In Fig. \ref{r_accEL}, we have plotted the actual and the Eddington accretion rates for six models; we also give the length of time each system is undergoing Eddington-limited accretion in the top left corner of each panel.
The unphysical model PNK$_\text{f}$r05t10 is undergoing Eddington-limited accretion for 11.7 per cent of the time, thus a (relatively) high percentage is a possible indicator of an unphysical model.  However, the unphysical Model PNK$_\text{f}$r02t05 undergoes Eddington-limited accretion 6.3 per cent of the time, which is higher than the per cent of time the physical Model DQM$_\text{f}$ is undergoing Eddington-limited accretion. While there is a correlation between the amount of time spent in Eddington-limited accretion and the final black hole mass for the PNK models, this is not true when comparing across different models. Further, these numbers indicate that relying upon the $M_\text{BH}$--$\sigma$ relationship appears to place the amount of time in Eddington-limited accretion in a comparatively narrow band. It is thus difficult for us to draw detailed conclusions about using the amount of time spent in Eddington-limited accretion as a strong discriminant between the physical or unphysical nature of models.

Ultimately, the amount of time accreting at the Eddington limit is very
closely related to the duty cycle of the quasar. In the context of
models such as \citet{SB92}, where the population of SMBH is
predicted as a function of luminosity (or mass) and time, the duty cycle
directly impacts the global population because the luminosity function
is a product of the number density, mass and duty cycle. Models of this
type have been extensively developed by Shankar (e.g. \citealp{SWM13}; \citealp{SCMFW10}) and show that
while Eddington accretion is likely to be more common at lower masses
($M<10^{7.5}$ M$_{\astrosun}$), once the SMBH reaches 10$^8$ M$_{\astrosun}$, matching the
luminosity function requires that growth at the Eddington rate must be
considerably less frequent or radiative efficiencies must be low. While obviously
not directly equivalent, matching the $M_\text{BH}$--$\sigma$ relationship appears to 
place a very similar constraint on the allowed duty cycle.

Lastly, the gas discs in the simulation are quite stable and in the absence of any additional supply of infalling gas actually decrease their mass during the simulation by 63 per cent prior to the start of the main merger.  The stability of the discs is a product both of our choosing a comparatively stable initial configuration and also that the SN feedback routine keeps the gas comparatively well supported against cold collapse. Thus the discs do not fuel the black hole through large scale fragmentation due to them becoming unstable.

%-----------------------------------------------------------------------
\subsection{Parameter sensitivity: $R_\text{acc}$}
\label{psr}

\begin{figure}
\begin{center}
\includegraphics[width=1.0\columnwidth]{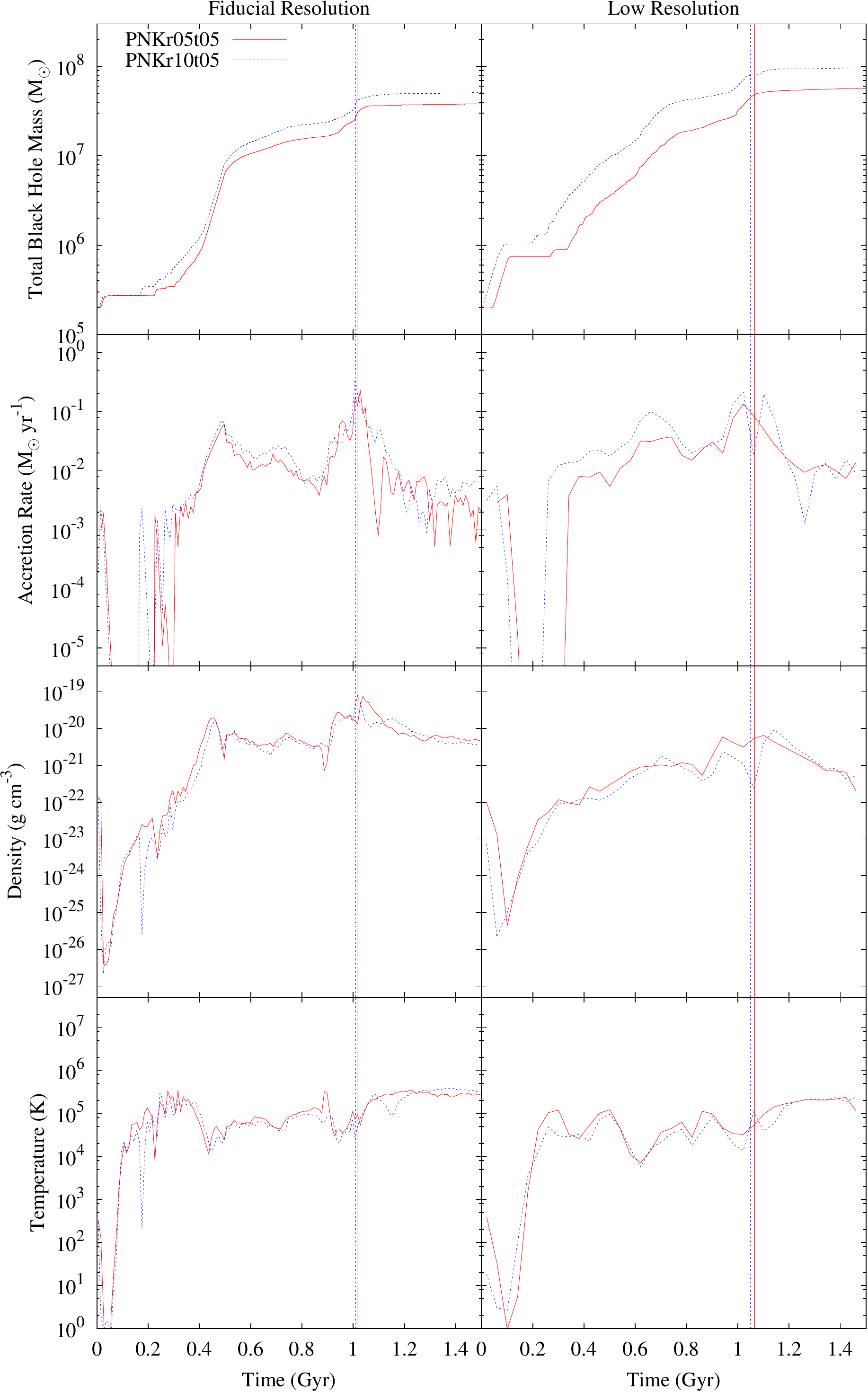}
\end{center}
\caption{The total black hole mass, accretion rates on to the black hole, gas density and gas temperature within $r_\text{inf}$ of the ADP, as in Fig. \ref{r_prop_ge}; we fix $t_\text{visc} = 5$ Myr and vary $R_\text{acc}$.} \label{r_prop_racc}
\end{figure}

We have four models (two models each of fiducial and low resolution) with $t_\text{visc} = 5$ Myr that produce physically plausible results.  In Fig. \ref{r_prop_racc}, we have plotted total black hole mass, the accretion rates on to the black hole, and the gas density and gas temperature within $r_\text{inf}$ of the ADP.
As expected, the first accretion event occurs in Model PNKr10t05, and Model PNKr10t05 has a slightly higher accretion rate for most of the simulation.  By increasing the radius of accretion by a factor of two (hence the volume in which a particle can be accreted, $V_\text{acc}$, by a factor of eight), there is only a factor of 1.35 and 1.70 increase in final total black hole mass for fiducial and low resolutions, respectively.  The gas density and temperature behaviour within $r_\text{inf}$ are very similar at both resolutions.  The value of $r_\text{inf}$ has no explicit dependence on $R_\text{acc}$ and all models produce the same qualitative behaviour for $r_\text{inf}$, which settles to $\sim 2h_\text{min}$ between first periapsis and apoapsis.  We find that $R_\text{acc}$ has negligible influence on $r_\text{inf}$ and a minimal impact on the final black hole masses.  Thus, we conclude that, as long as the parameters are in the physically acceptable range, results are not overly sensitive to the exact value of $R_\text{acc}$.

%-----------------------------------------------------------------------
\subsection{Parameter sensitivity: $t_\text{visc}$}
\label{pst}

\begin{figure}
\begin{center}
\includegraphics[width=1.0\columnwidth]{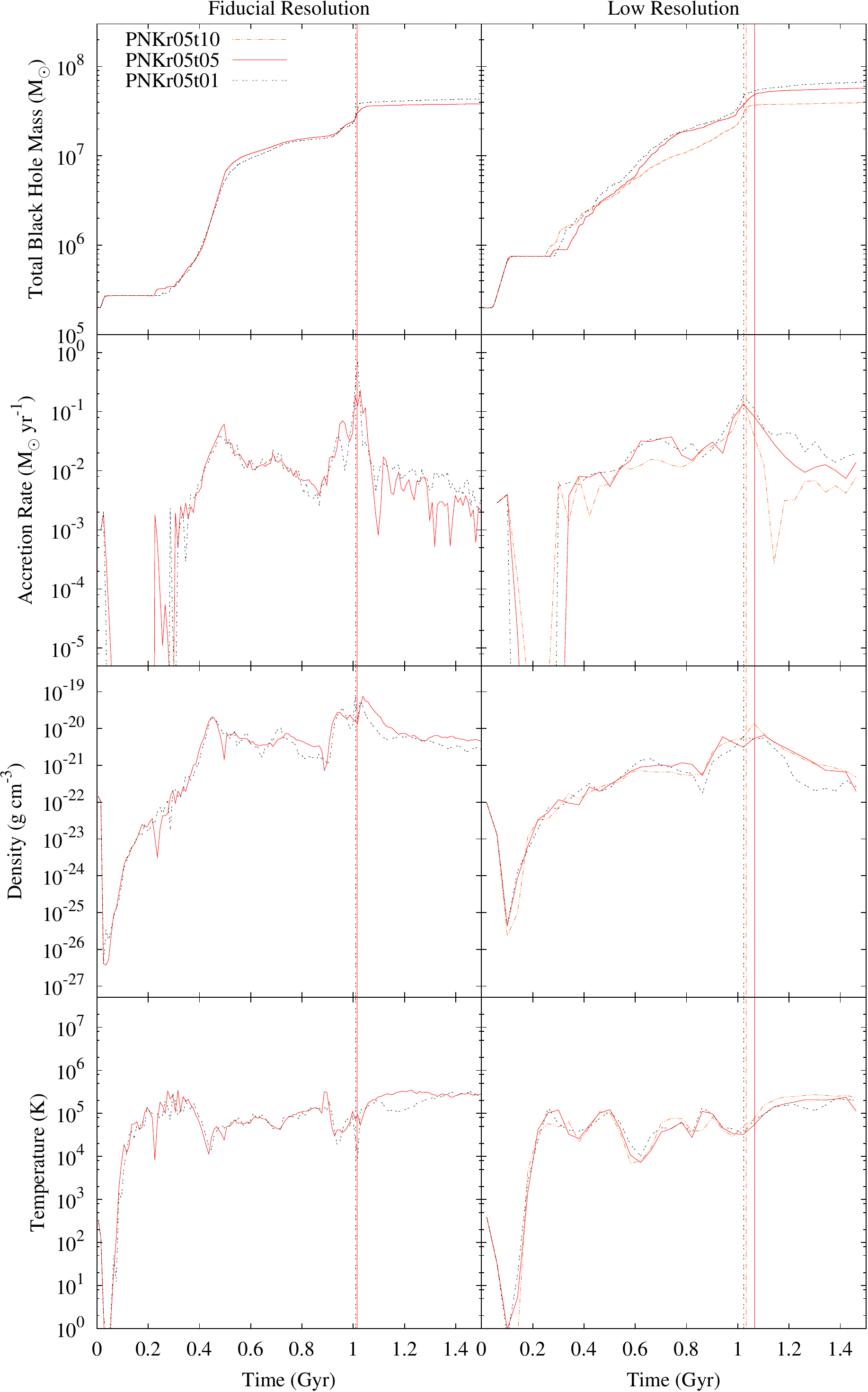}
\end{center}
\caption{The total black hole mass, accretion rates on to the black hole, gas density and gas temperature within $r_\text{inf}$ of the ADP, as in Fig. \ref{r_prop_ge}; we fix $R_\text{acc} = 0.05h_\text{min}$ and vary $t_\text{visc}$.}\label{r_prop_tvisc}    
\end{figure}

We have five models with $R_\text{acc} = 0.05h_\text{min}$ (two at the fiducial resolution and three at low resolution).  In Fig. \ref{r_prop_tvisc}, we have again plotted total black hole mass, the  accretion rates on to the black hole, and the gas density and gas temperature within $r_\text{inf}$ of the ADP.
The total black hole masses in the fiducial resolution simulations are similar throughout the simulations, and their final masses differ by 13 per cent.  At low resolution, the total black hole masses for models PNK$_\text{l}$r05t05 and PNK$_\text{l}$r05t01 match within 18 per cent at the end of the simulation; if we include PNK$_\text{l}$r05t10 (a model that produces unphysical behaviour at the fiducial resolution), then the low resolution black hole masses differ at most by a factor of 1.68.  

\begin{figure}
\begin{center}
\includegraphics[width=1.0\columnwidth]{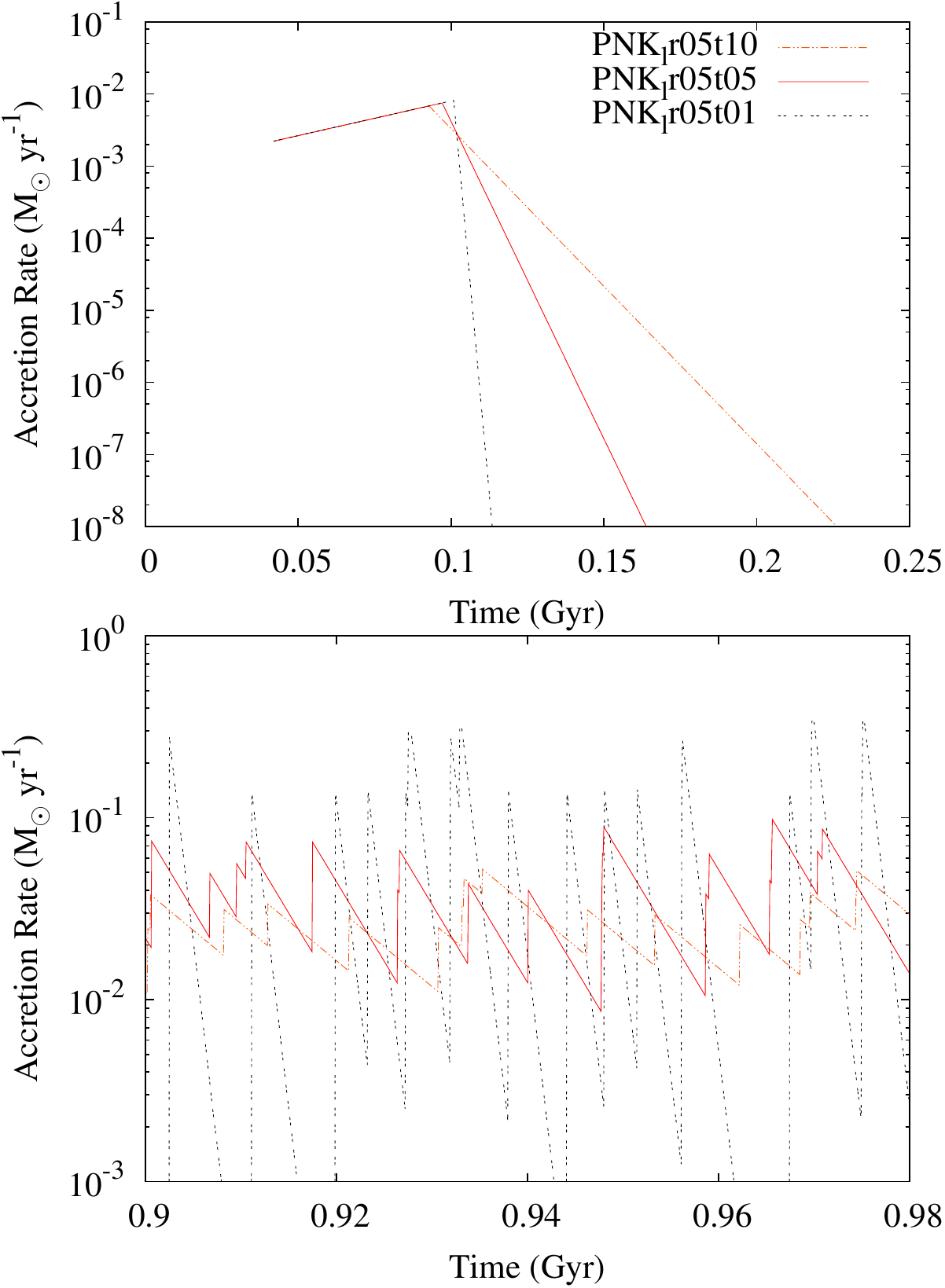}
\end{center}
\caption{The gas accretion rate on to one black hole for three low resolution models, holding $R_\text{acc} = 0.05h_\text{min}$.  We show the initial accretion event (top) and the accretion just prior to core merger (bottom).}\label{r_prop_acc2}
\end{figure}

In Fig. \ref{r_prop_acc2}, we plot two segments of the accretion rate for one black hole in our three low resolution models with $R_\text{acc} = 0.05h_\text{min}$.
As expected, the first accretion event happens simultaneously in each model (top panel).  This is followed by a period of Eddington accretion, which stops first for $t_\text{visc} = 10$ Myr and last for $t_\text{visc} = 1$ Myr depending upon when the $M_\text{disc}/t_\text{visc}$ rate falls below the Eddington rate.  Once the Eddington accretion has stopped, the rate of feedback (where $\dot{p} \propto \dot{M}$) differs for each model, but at least at this early time the accretion rate on to the black hole can still be calculated: an exponentially decaying rate with the time dependence given by $t/t_\text{visc}$.  Although PNKr05t01 accretes at the Eddington limit very slightly longer than the other two models, the faster decrease in the accretion rate means that feedback effectively stops sooner, thus giving particles more time to slow down and fall within $R_\text{acc}$. 

At late times, each accretion event does not add enough mass to the disc to permit Eddington accretion (bottom panel); however, there is a local accretion peak followed by exponential decay.  As previously discussed, the faster decay of $t_\text{visc} = 1$ Myr allows for rapid accretion events on to the disc; this creates large variances in the accretion rates, spanning a few orders of magnitude.  The slower decay of $t_\text{visc} = 10$ Myr yields a moderated accretion rate, thus there is only a factor of a few between the local peak accretion rate and the minimum rate prior to the next accretion event.  The time-average accretion rate yields a higher accretion rate for the lower viscous time-scales, resulting in more feedback being returned to the gas which can, at least temporarily, expel considerable amounts of gas from the system.  Summarily, we find a general trend to more massive black holes with decreasing $t_\text{visc}$.

The torques from the interacting galaxies modify this argument by introducing disc instabilities  Although the feedback rate of the $t_\text{visc} = 10$ Myr models can hinder gas from accreting on to the disc during a quiescent phase, there is not enough cumulative feedback to prevent a large gas flow on to the disc produced by the disc instabilities.  This results in a large accretion epoch, followed by powerful outbursts of feedback energy.  Thus, the tidal torques at first periapsis and core merger are strong enough to over come the feedback and cause a catastrophic (or near catastrophic) outburst, as in Model PNK$_\text{f}$r05t10.

Similar to our discussion in section \ref{psr}, we find quantitatively similar gas behaviour within $r_\text{inf}$, which is a result of the $r_\text{inf}$'s being similar across all models.  As with $R_\text{acc}$, we conclude that, as long as $t_\text{visc}$ lies with the physically acceptable range, then the simulations are comparatively insensitive to its exact value (at least to within 1/2 an order of magnitude at the fiducial resolution).

%-----------------------------------------------------------------------
\subsection{Resolution sensitivity}
\label{rs}

\begin{figure*}
\begin{center}
\includegraphics[width=1.0\textwidth]{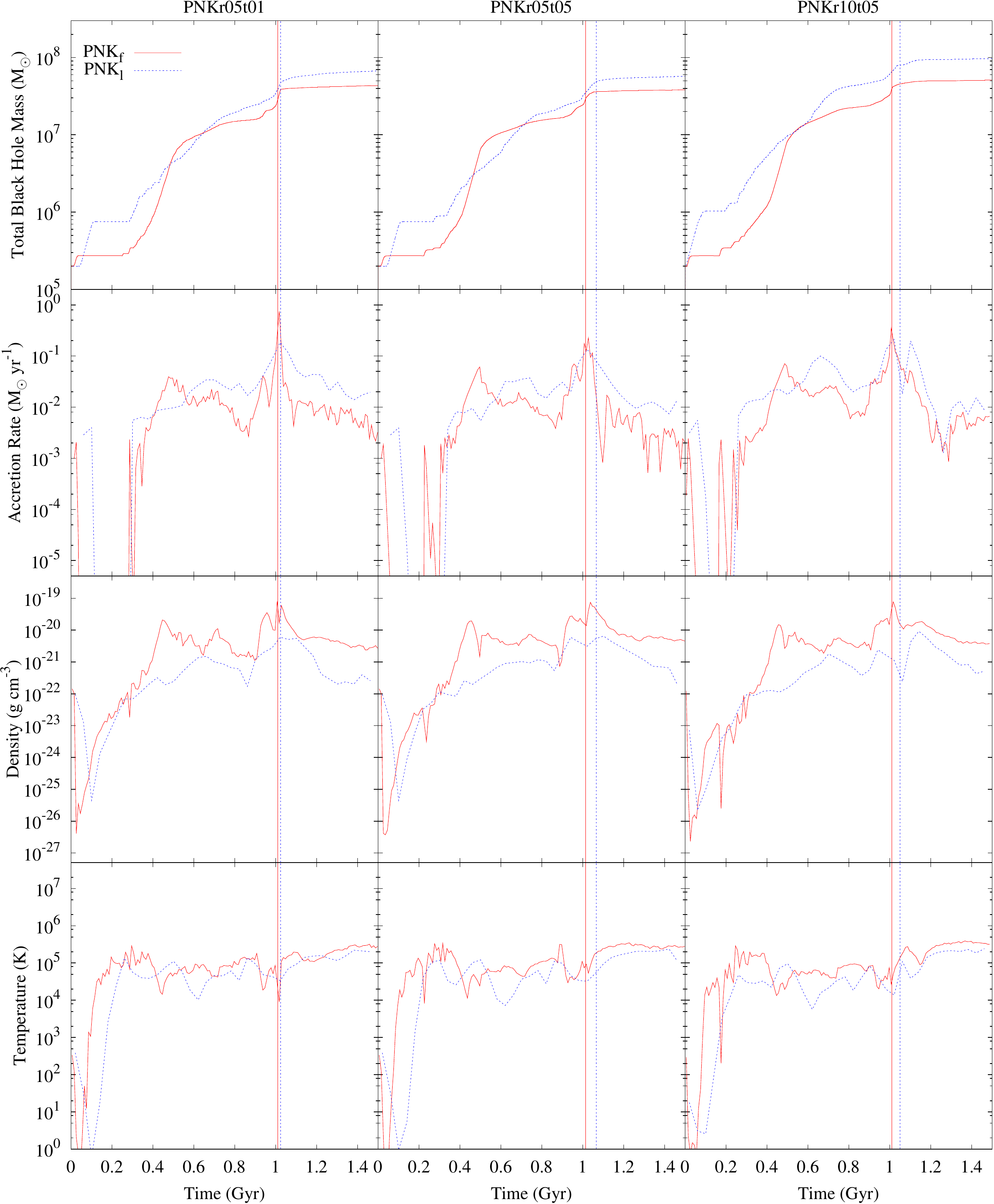}
\end{center}
\caption{The total black hole mass, accretion rates on to the black hole, gas density and gas temperature within $r_\text{inf}$ of the ADP, as in Fig. \ref{r_prop_ge}.  Each column plots data for a different set of $R_\text{acc}$ and $t_\text{visc}$.}\label{r_prop_res}    
\end{figure*}

We have three models that are deemed to have physically plausible results at both resolutions.  In Fig. \ref{r_prop_res}, we have plotted total black hole mass, the accretion rates on to the black hole, and the gas density and gas temperature within $r_\text{inf}$ of the ADP.
When we directly compare two models at different resolutions, there are significantly more differences in accretion and evolution histories than we do for when the model parameters are changed (at least for the acceptable range).  As would be expected, in all cases the fiducial resolution model accretes the first particle, while the lower resolution simulations produce a larger jump in the accretion rate at early times due to the particles in those simulations being more massive.  By the time the galaxies reach first periapsis at 166 Myr, the low resolution gas has a greater kinetic feedback obstacle to overcome before it can pass within $R_\text{acc}$ of the ADP.  Thus, the first low resolution accretion event immediately begins to moderate the accretion on to the disc, and this moderation continues to persist after first periapsis.

Because it has a smaller kinetic feedback obstacle to overcome, the fiducial resolution gas can more easily fall within $R_\text{acc}$ of the ADP after first periapsis, leading to a large and essentially unmoderated accretion on to the disc.  As the gas accretes from the disc on to the black hole, it modifies the environment to suppress further accretion events on to the disc; however, the disc remains massive from the previous accretion episode.  Thus, there is a decrease in gas density in the core, but the accretion on to the black hole continues as the gas in the disc is continually transferred to the black hole.  This major epoch of accretion on to the disc after first periapsis results in a steeper black hole growth between first periapsis and apoapsis for the fiducial resolution models. 

For all three models, the fiducial resolution models have the expected higher core gas densities. Since the feedback is being returned kinetically, shock heating and star formation are the primary heating mechanisms, and both scale with resolution in turn producing similar core temperatures.

Thus, based upon the two resolutions we test, we can conclude that resolution has a greater impact on the results than the values of the free parameters, $R_\text{acc}$ and $t_\text{visc}$.  However, these differences are predictable since there are more epochs of discrete behaviour in the fiducial resolution models than in the low resolution models, allowing for a more continuous modelling of the evolution.  Also, each fiducial--low resolution pair is more similar to one another than to Models SDH or DQM of the same resolution, indicating that the model can be distinguished from other models even at low resolution.

%-----------------------------------------------------------------------
\subsection{Final states}
\label{frms}
\subsubsection{Stellar remnant}
\begin{figure}
\begin{center}
\includegraphics[width=1.0\columnwidth]{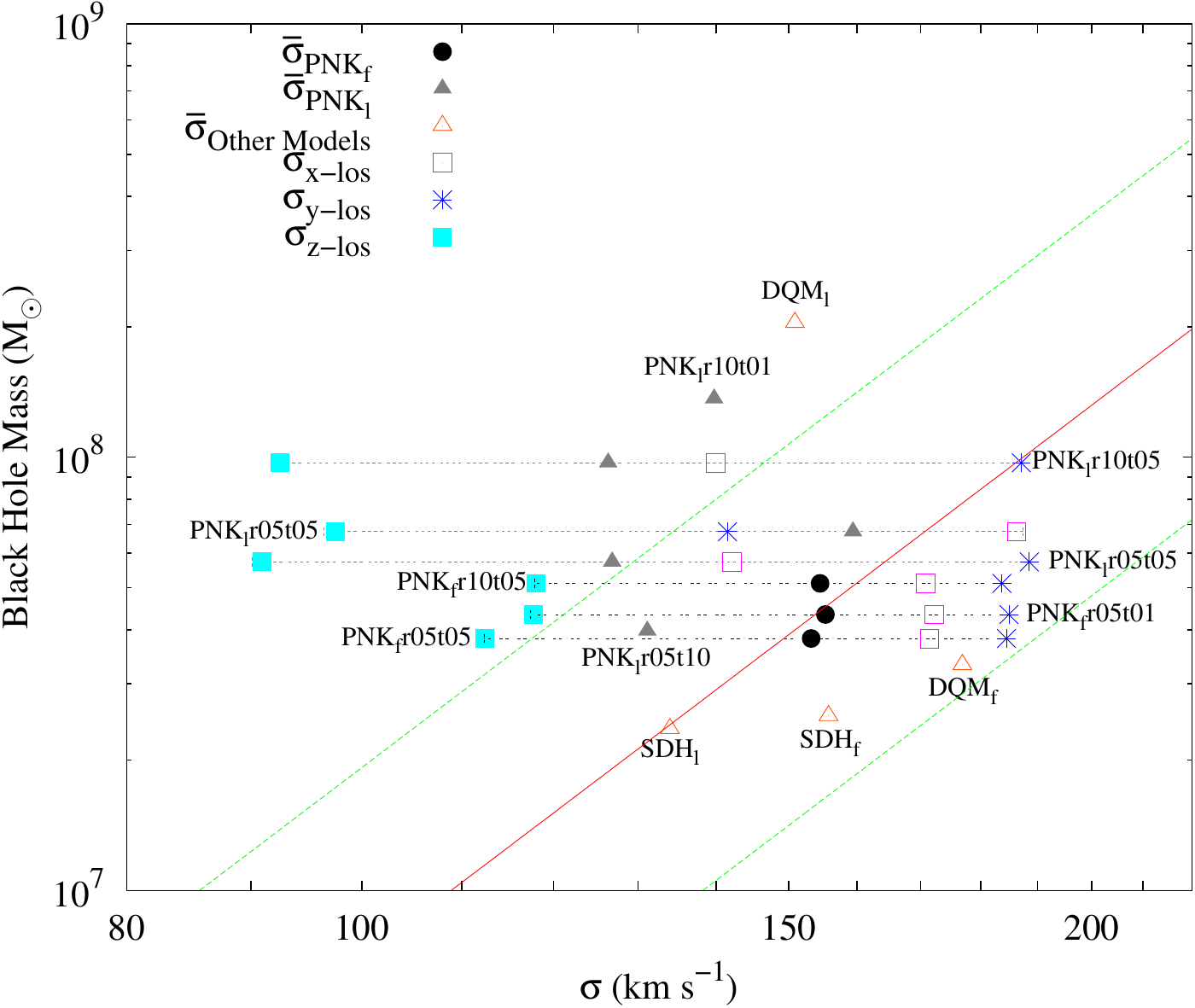}
\end{center}
\caption{Our numerical $M_\text{BH}$--$\sigma$ relation, along with the observed relation (red solid line) and the one-sigma scatter (green dashed line) from \citet{Getal09}.  For six selected models, the solid dot/triangle represents the average $\sigma$ of 1000 random lines of sight, the horizontal bars represent the range of all calculated $\sigma$'s, and the remaining three symbols on the horizontal line represent $\sigma$ taken along the $\pm x-$, $\pm y-$ and $\pm z-$lines of sight.  For the remainder of our physical PNK models, we have only plotted the average $\sigma$ of 1000 random lines of sight.  We have explicitly labelled which points/set of points corresponds to which models.}\label{msigma} 
\end{figure}

\begin{figure*}
\begin{center}
\includegraphics[width=0.65\textwidth]{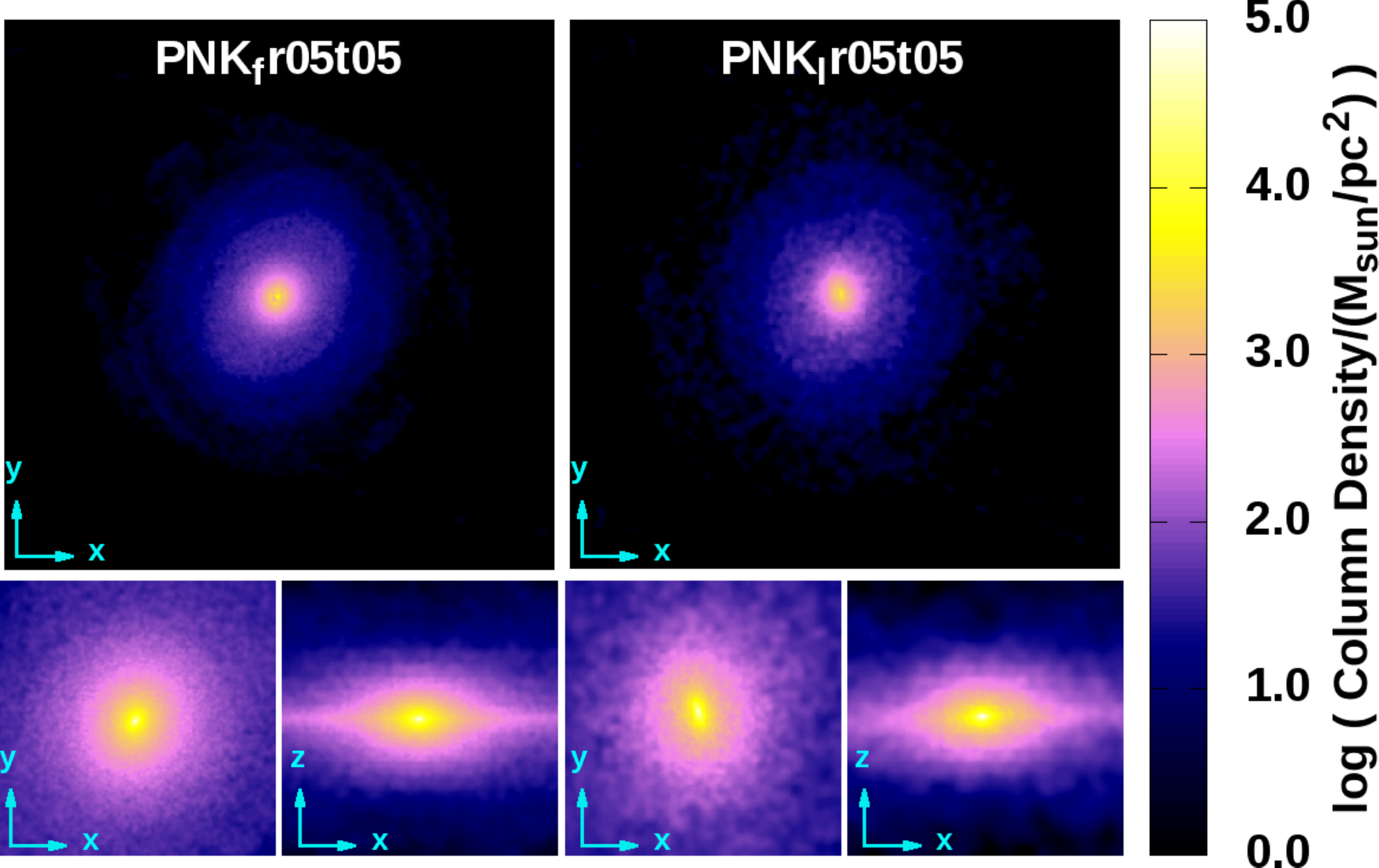}
\end{center}
\caption{Stellar column density of the remnants of PNK$_\text{f}$r05t05 (left) and PNK$_\text{l}$r05t05 (right), taken at 1.5 Gyr.  All stellar remnants (including Models DQM and SDH) of each resolution are similar, with the low resolution remnants yielding more elliptical bulges than their fiducial resolution counterparts. \emph{Top Row}: Face-on view with each frame measuring 100 kpc per side, with image resolution of 195 pc/pixel (left) and 390 pc/pixel (right). \emph{Bottom}: Face-on and edge-on view of both models, with each frame measuring 20 kpc per side, and image resolution of 39 pc/pixel (78 pc/pixel) for the fiducial (low) resolution models.}\label{f_finalS}  
\end{figure*}

A common test of numerical accretion and feedback algorithms is the $M_\text{BH}$--$\sigma$ relation.  For all of our physical models, we have calculated the stellar velocity dispersion around every black hole using the method described in \citet{DQM11}.  These velocity dispersions are averaged over 1000 random lines of sight, and are plotted on the $M_\text{BH}$--$\sigma$ relation in Fig. \ref{msigma}.
For the six models in Fig. \ref{r_prop_res}, we also plot the full range of velocity dispersions, as well as those taken preferentially along the $\pm x$-, $\pm y$- and $\pm z$-lines of sight.  The large range of $\sigma$ is a result of the highly triaxial stellar remnant; see Fig. \ref{f_finalS}, where we have plotted a fiducial and low resolution stellar remnant.  The fiducial resolution models have average values very near the observed $M_\text{BH}$--$\sigma$ relation, and a range that nearly stays within the one-sigma scatter.  The low resolution stellar remnants are more elliptical than their fiducial resolution counterparts, thus they have a greater range of $\sigma$'s. While some of the average $\sigma$'s fall outside of the one-sigma range, they all obtain $\sigma$'s near the observed $M_\text{BH}$--$\sigma$ relation if the line of sight is near the plane of the disc.  To verify these results, we have recalculated the velocity dispersions of the fiducial resolution models based upon the gravitational potential of the black hole, and found these values to be consistent within five per cent of the average values reported above.  

%-----------------------------------------------------------------------
\subsubsection{Gas properties}
\begin{figure}
\begin{center}
\includegraphics[width=0.96\columnwidth]{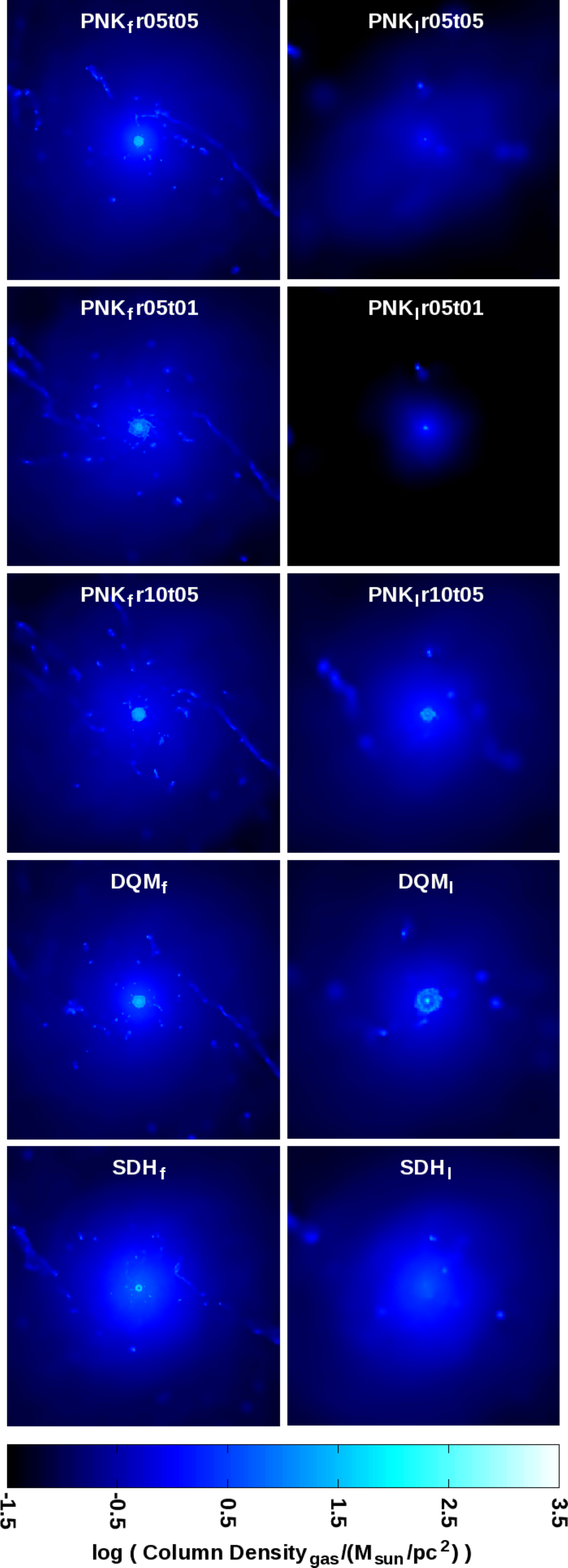}
\end{center}
\caption{Gas column density of ten remnant, taken at 1.5 Gyr.  Each frame measures 100 kpc per side, with image resolution of 195 pc/pixel (390 pc/pixel) for the fiducial (low) resolution models.}\label{f_finalDbig}  
\end{figure}

\begin{figure*}
\begin{center}
\includegraphics[width=0.85\textwidth]{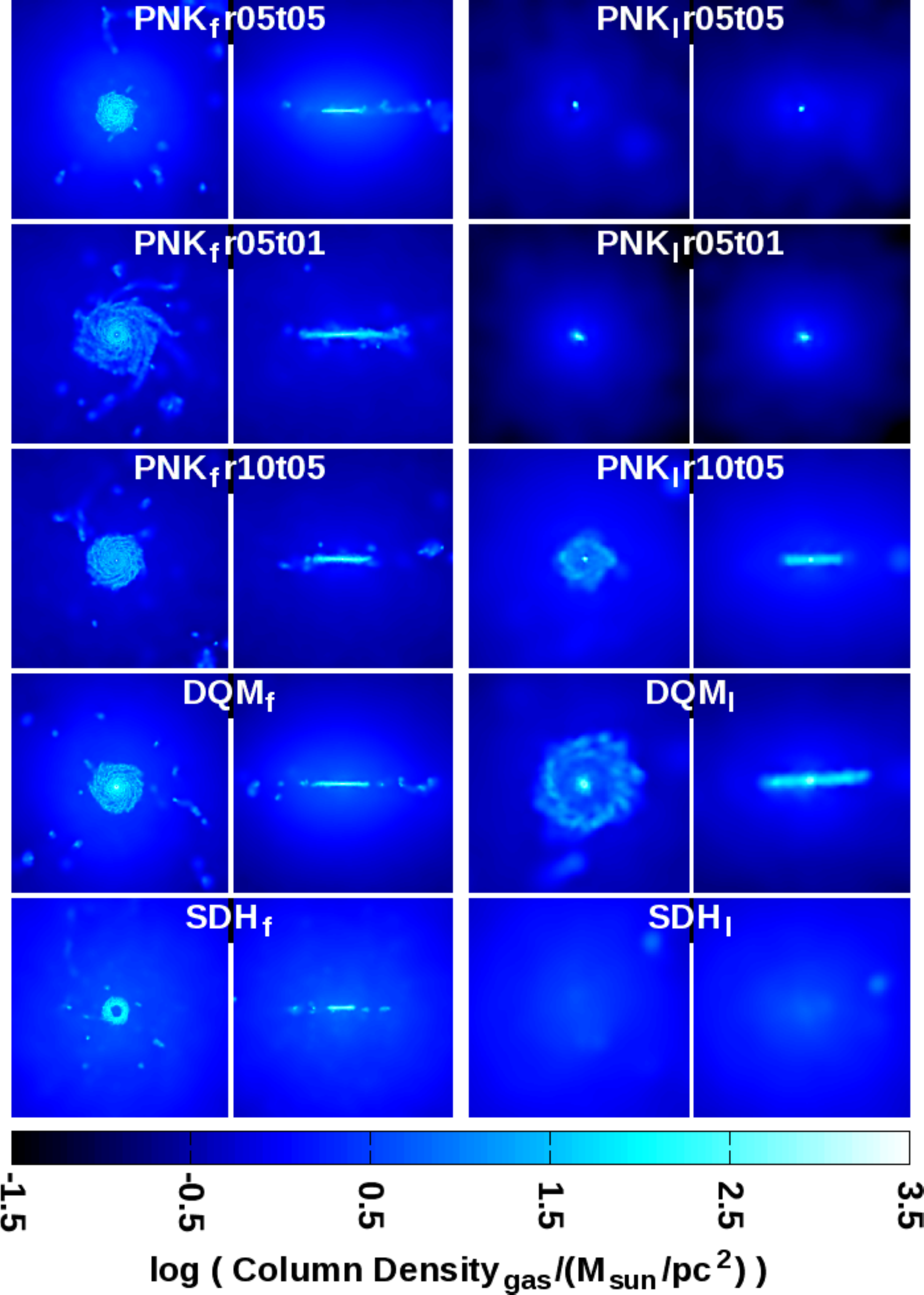}
\end{center}
\caption{Gas column density of ten remnant, taken at 1.5 Gyr.  Each pair contains a face-on and edge-on view of the central 20 kpc.  The image resolutions are 39 pc/pixel (78 pc/pixel) for the fiducial (low) resolution models.}\label{f_finalDsmall}  
\end{figure*}

In Figs. \ref{f_finalDbig} and \ref{f_finalDsmall}, we plot the gas column density of ten remnants at 1.5 Gyr.  These show, respectively, the inner 100 and 20 kpc of six PNK models (three models as each resolution) and Models DQM and SDH.  
As discussed in sections \ref{psr} and \ref{pst}, there is much qualitative similarity amongst the three fiducial remnants of the PNK models, as well as Models DQM$_\text{f}$ and SDH$_\text{f}$.  In all cases, there is a condensing gas cloud with a reformed disc.  

\begin{figure}
\begin{center}
\includegraphics[width=1.0\columnwidth]{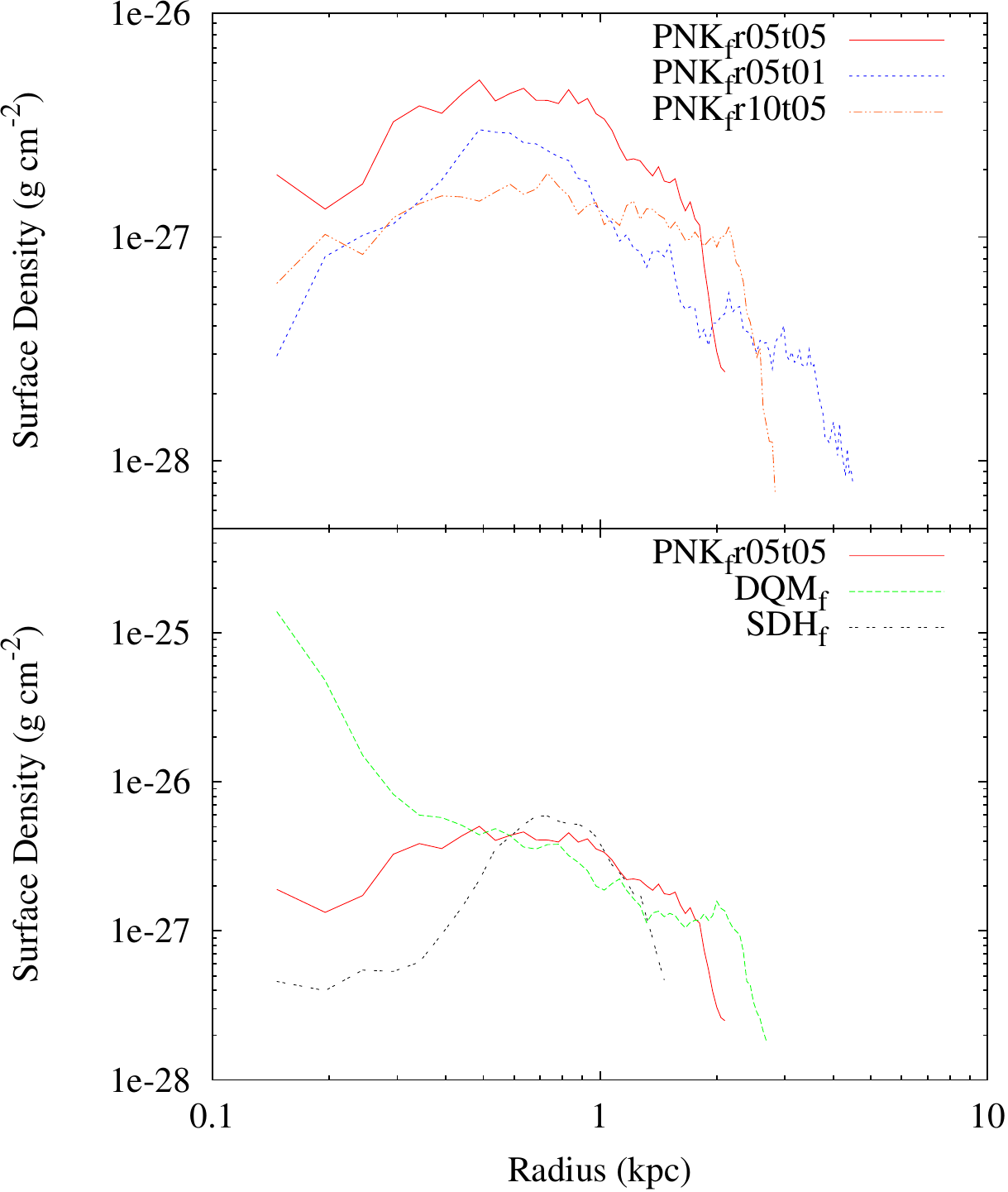}
\end{center}
\caption{Gas surface density profile, averaged over all azimuthal angles, for the reformed disc of three fiducial resolution remnants.  The profile is truncated at the edge of the disc, and plotted in bins of 49 pc.}\label{f_finalGSD}  
\end{figure}

The radius, surface density profiles and total gas mass of the reformed discs in the PNK models are only slightly dependent on the free parameters; the surface density profile is plotted in the top panel of Fig. \ref{f_finalGSD} and the gas disc masses are given in Table \ref{discmass}.
\begin{center}
\begin{table}
{\small
\hfill{}
\begin{tabular}{l|c }
    \hline
Model  & Disc mass ($10^8$ M$_{\astrosun}$) \\
\hline
\hline
PNK$_\text{f}$r05t05 & 4.18 \\
PNK$_\text{f}$r05t01 & 3.93 \\
PNK$_\text{f}$r10t05 & 3.31 \\
DQM$_\text{f}$       & 6.42 \\
SDH$_\text{f}$       & 2.70 \\
\hline
\end{tabular}}
\hfill{}
\caption{The gas mass of the reformed discs at 1.5 Gyr for five fiducial resolution models.}  \label{discmass} 
\end{table}
\end{center}

After core merger, PNK$_\text{f}$r10t05 has a slightly larger accretion rate than PNK$_\text{f}$r05t05 due to its larger accretion radius, this leads to more feedback energy and a slightly extended disc with lower surface density.  Model PNK$_\text{f}$r05t01 has a higher accretion rate at core merger than PNK$_\text{f}$r05t05, which results in a small outburst at 1 Gyr, expelling some gas from the system. By 1.5 Gyr, the remaining gas is less bound leading to the slightly larger radius and lower surface density.

The fiducial resolution PNK models have different surface density profiles than Models DQM$_\text{f}$ and SDH$_\text{f}$; see the bottom panel of Fig. \ref{f_finalGSD} and note the difference range on the vertical axis.  The disc in Model DQM$_\text{f}$ has a dense core and a moderately dense torus (resulting in a high-mass disc), whereas the profile for Model SDH$_\text{f}$ is a torus due to low level feedback activity carving out a region of the disc. 

The three fiducial resolution PNK remnants are qualitatively more similar to one another than to their low resolution counterparts.  Both PNK$_\text{l}$r05t01 and PNK$_\text{l}$r05t05 underwent a major outburst starting at 1.05 Gyr, blowing away much of the gas and preventing the reformation of the gas disc.  In the subsequent few 100 Myrs, the gas begins to recondense into the cloud presented here.  In PNK$_\text{l}$r10t05, this major outburst never occurs, thus allowing the disc to reform.

%******************************************************************************************************************************************************************************
%The paper - Conclusion
\section{Conclusion}
\label{conclusion}
We have implemented the accretion disc particle (ADP) method of \citet{PNK11} into a major merger simulation of two Milky Way-sized galaxies.  We ran five fiducial resolution simulations and twelve low resolution simulations varying the free parameters $R_\text{acc}$ and $t_\text{visc}$.  Our primary conclusions 
are as follows:
\begin{enumerate}

\item 
For accretion radii that are too small (i.e. $R_\text{acc} = 0.02h_\text{min}$), the final black hole mass is far smaller than predicted from the $M_\text{BH}$--$\sigma$ relationship.  Thus all of our $R_\text{acc} = 0.02h_\text{min}$ models are classified as unphysical. 

\item
For accretion radii that are too large (i.e. $R_\text{acc} = 0.20h_\text{min}$), the accretion rate on to the accretion disc (hence on to the black hole) is unreasonably large.  The resulting feedback is enough to catastrophically disrupt the system.  We thus classify all of our $R_\text{acc} = 0.20h_\text{min}$ models as unphysical.

\item
For large viscous time-scales (i.e. $t_\text{visc} = 10$ Myr), feedback from an accretion event persists long enough to hinder secular accretion.  Tidal forces from the interacting galaxies can overcome the low amount of feedback and funnel considerable amounts of gas on to the accretion disc.  Depending on resolution and accretion radius, this short accretion epoch can be large enough such that its resulting feedback energy can catastrophically disrupt the system.  Thus a few of our $t_\text{visc} = 10$ Myr models are classified as unphysical.  

\item
The values of $R_\text{acc}$ and $t_\text{visc}$, assuming they were in the allowed parameter space, had minimal affect on the gas properties within $r_\text{inf}$ of the black hole.  This result was expected since $r_\text{inf}$ had no explicit dependence on $R_\text{acc}$ or $t_\text{visc}$.

\item 
The exact value of $R_\text{acc}$ has only a minimal affect on the resulting system, assuming that it is in the allowed parameter space.  By doubling the accretion radius from $0.05h_\text{min}$ to $0.10h_\text{min}$, the final black hole mass increases by only a factor of 1.35 (1.70) for our fiducial (low) resolution simulations.

\item 
As we decrease the value of $t_\text{visc}$ the final black hole mass increases; the final range of black hole masses spans a factor of 1.13 (1.68) for our fiducial (low) resolution models.  Thus, the exact value of $t_\text{visc}$ has only a minimal affect on the resulting (physical) system.  

\item 
Decreasing the resolution increases the final black hole mass; for any given fiducial--low resolution pair, the final black hole mass differs by at most a factor of 1.90.  The fiducial resolution models experience steeper black hole growth between first periapsis and apoapsis, and the lower resolution models are more prone to major outburst events shortly after core merger.  Given the parameters tested here, resolution has the largest impact on the outcome of the model.   We understand that these differences are unavoidable and are a result of a single accretion event in the low resolution models returning 7.6 times more energy to the gas than the fiducial resolution models.

\end{enumerate}

In AGN feedback models where the accretion rate is dependent on the gas properties near the black hole, the accretion and energy feedback rates are never zero, and span only a few orders of magnitude; in our simulations, the accretion and feedback begins immediately.  In Model DQM, an initially high accretion rate leads to an immediate and dramatic modification of the black hole's environment.  In Model SDH$_\text{f}$, the accretion rate does not become significant 
until apoapsis, at which point the black hole undergoes a major accretion epoch, with its mass increasing by a factor of 20.5 in 470 Myr.

With the ADP model, the initial accretion rate is zero or very small.  The minimal modification of the ADP's environment by first periapsis allows a large accretion event on to the disc, which results in a large accretion rate on to the black hole, leading to a very rapid black hole growth; in the case of Model PNK$_\text{f}$r05t05, the black hole mass increases by a factor of 19.6 in 200 Myr.  Although these three models, Models SDH, DQM and PNK, have very different algorithms and different evolution histories, there are similar remnants in all three cases.

The ADP algorithm decouples the accretion rate on to the black hole from the gas properties around it and allows a high accretion rate even after feedback has prevented additional accretion events on to the disc. Yet to have confidence in this conceptual model a precise observational strategy for determining parameters is necessary (essentially looking at the correlation, or lack of, between activity and the nuclear gas environment). Recent observational work \citep{WHC10} has unveiled how star formation activity and black hole growth appear tied together in spheroids. These results undoubtedly give useful hints on how black hole growth proceeds outside of rapid accretion phases, but fueling of black hole via cold gas accretion remains the biggest uncertainty in our models currently. To date, studies of nuclear CO morphology such as the NUGA project (e.g. \citealt{NUGA4}; \citealt{NUGA15}) have found no obvious morphological links between local AGN activity and mid-scale CO morphology. But such studies are obviously limited by resolution concerns, as well as probing an entirely different AGN luminosity regime. However, ALMA in its full configuration with the largest baselines will provide resolutions that are sufficient to study the nuclear regions the nearest active quasar systems. The prospect of examining morphology dependence in these systems is immensely exciting and should provide great insight into their evolution and how they can best be modelled.

%******************************************************************************************************************************************************************************
\section*{Acknowledgments}

We thank the anonymous referee for helpful comments that improved the content and clarity of the manuscript.  We thank Larry Widrow for providing his NFW halo generator.  JW is supported by NSERC and Saint Mary's University.  RJT is supported by a Discovery Grant from NSERC, the Canada Foundation for Innovation, the Nova Scotia Research and Innovation Trust and the Canada Research Chairs Program. Simulations were run on the CFI-NSRIT funded {\em St Mary's Computational Astrophysics Laboratory}.
%******************************************************************************************************************************************************************************
\appendix
\section{Galaxy models}
\label{simsGM}
We construct our model galaxy to have similar mass components to the galaxies presented in \citet{SDH05}.  We begin by using the GalactICs package (\citealp{KD95}; \citealp{WD05}; \citealp{WPD08}) to create a Milky Way-sized galaxy that consists of a stellar bulge, stellar disc, and a dark matter halo; this is done through an iterative process to produce a self consistent system.  The free parameters, and the values we chose, are listed in Table \ref{galprops}.
\begin{center}
\begin{table}
{\small
\hfill{}
\begin{tabular}{l l l}
\hline
Component             & Parameter    	    & Value     	   \\
\hline
\hline
Bulge                 & $\sigma_\text{b}$   & 292   km s$^{-1}$    \\
                      & $R_\text{e}$        &   0.7 1kpc 	   \\
                      & $n$     	    &   1.1     	   \\
\hline
Disc                  & $R_\text{d}$        &   2.46 kpc 	   \\
                      & $z_\text{d}$        &   0.49 kpc 	   \\
                      & $R_\text{trunc}$    &  30    kpc    	   \\
                      & $z_\text{trunc}$    &   1    kpc    	   \\
                      & ${\sigma_\text{R}}_0$ & 119    km s$^{-1}$   \\
\hline
Dark Matter Halo      & $a_\text{h}$        &  13.6  kpc  	   \\
                      & $r_\text{h}$        & 275    kpc    	   \\
                      & $\delta r_\text{h}$ &  25    kpc    	   \\
                      & $\sigma_\text{h}$   & 330    km s$^{-1}$   \\
                      & $\gamma$     	    &   0.81    	   \\
\hline
metallicity           & $Z$                 & 0.05 Z$_{\astrosun}$ \\
mean molecular weight & $\mu$               & 0.6       	   \\
\hline
\end{tabular}}
\hfill{}
\caption{The chosen parameters for our model galaxies.  All parameters are defined in Appendix \ref{simsGM}.}
\label{galprops} 
\end{table}
\end{center}

The dark matter halo profile follows
\begin{equation}
\label{GIChalo}
\tilde{\rho}_\text{h} = \frac{2^{2-\gamma}\sigma^2_\text{h}}{4\pi a^2_\text{h}} \frac{1}{(r/a_\text{h})^\gamma (1+r/a_\text{h})^{3-\gamma}} C(r;r_\text{h},\delta r_\text{h}),
\end{equation}
where $a_\text{h}$ is the halo scale length, $r_\text{h}$ is the cutoff radius, $\gamma$ is the central cusp strength and $\sigma_\text{h}$ is a (line of sight) velocity scale that sets the mass of the halo.  The truncation function, 
$C(r;r_\text{h},\delta r_\text{h}) = \frac{1}{2}\text{erfc}\left( \frac{r-r_\text{h}}{\sqrt{2}\delta r_\text{h}} \right)$, smoothly goes from one to zero at $r = r_\text{h}$ over width $\delta r_\text{h}$.  

The stellar disc has a truncated density profile that falls off approximately exponentially in $R$ and follows sech$^2$ in $z$; the disc has radial and vertical scale heights $R_\text{d}$ and $z_\text{d}$ and truncation distances $R_\text{trunc}$ and $z_\text{trunc}$.  The radial velocity dispersion is given by $\sigma_\text{R}^2(R) = {\sigma^2_\text{R}}_0 e^{-R/R_\sigma}$, where ${\sigma_\text{R}}_0$ is the central velocity dispersion and $R_\sigma = R_\text{d}$ for simplicity.  

The stellar bulge density profile is given by
\begin{equation}
\label{GICbulge}
\tilde{\rho}_\text{b}(r) = \rho_\text{b} \left( \frac{r}{R_\text{e}} \right)^{-p} \text{e}^{-b(r/R_\text{e})^{1/n}},
\end{equation}
which yields a S\'ersic law for the projected density profile if $p = 1 - 0.6097/n + 0.05563/n^2$, where $n$ is a free parameter.  The constant $\rho_\text{b}$ is defined using $\sigma_\text{b} \equiv \left\{ 4\pi n b^{n(p-2)}\Gamma \left[n(2-p)\right]R_\text{e}^2 \rho_\text{b}\right\}^{1/2}$, where $\sigma_\text{b}^2$ is the depth of the gravitational potential associated with the bulge and $R_\text{e}$ is the radial scale parameter.  The variable $b$ is adjusted such that $R_\text{e}$ encloses half the total projected light or mass.  

The galaxy is then modified in three ways.  First, ten per cent of the total stellar mass is converted into gas to create a gas disc.  The gas disc is the same as the stellar disc except that it has been reflected in the $x = y$ plane to avoid coincidence with the star particles.  The gas scale height is initially larger than physically motivated, however, cooling allows the gas to collapse into a thin disc within a few 10 Myr.  This vertical collapse produces a short transient evolution of the gas accompanied by a brief increase in the star formation rate.  At resolutions higher than presented here, this produces a strong ring-shaped shock which propagates outwards; one possible solution is to reduce the scale height of the gas disc, but studies in \citet{WT12} show that this is not necessary for the resolutions presented here.   

Second, a hot gas halo is added following the observationally motivated $\beta$-profile (e.g. \citealp{CF76}):
\begin{equation}
\label{betaprofile}
\rho(r) = \rho_0\left[1 + \left(\frac{r}{r_\text{c}}\right)^2\right]^{-\frac{3}{2}\beta},
\end{equation}
where $\rho_0$ is the central density, $r_\text{c}$ is the core radius, and $\beta$ is the outer slope parameter; we choose $r_\text{c} = 1.75$ kpc and $\beta = 2/3$ as done in \cite{MMSNC11}.  We set $\rho_0$ by choosing the mass of the hot gas halo within 40 kpc to be equal to two per cent of the total disc mass \citep{Retal09}.  To conserve total halo mass, we reduce the mass of the dark matter halo by the total mass of the hot gas halo.  By assuming isotropy and hydrostatic equilibrium, the temperature profile of the hot gas halo is given by \citep{KMWSM07}
\begin{equation}
\label{hghtemp}
T(r) = \frac{\mu m_\text{p}}{k_\text{B}} \frac{1}{\rho_\text{halo gas}(r)} \int_r^\infty  \rho_\text{halo gas}(r) \frac{GM(r)}{r^2}  \mathrm{d} r, 
\end{equation}
where $\mu$ is the mean molecular weight, $m_\text{p}$ is the proton mass, $k_\text{B}$ is the Boltzmann constants, and $M(r)$ is the total mass interior to $r$.  The halo gas is given an initial angular momentum which scales with the circular velocity, $j(R) \propto R v_\text{circ}(R)$, where $R$ is the distance from the spin axis of the galaxy \citep{MMSNC11}.  

Finally, a black hole (sink) particle is placed at the centre of the galaxy.  The galaxy has a total mass of $9.60\times 10^{11}$ M$_{\astrosun}$, 1 287 743 (168 351) particles for the fiducial (low) resolution simulations, and a Plummer softening length of $\epsilon_\text{Plummer} = 120$ pc ($\epsilon_\text{Plummer} = 300$ pc).

%******************************************************************************************************************************************************************************
%Bibliography
\bibliography{WTbib}

%******************************************************************************************************************************************************************************
%The End
\label{lastpage}
\end{document}